\documentclass{jfm_arxiv}
\newif\ifshowedits
\newif\ifshowdeletions

\usepackage{graphicx}
\usepackage{xcolor}
\usepackage{epstopdf,epsfig}
\usepackage{natbib}
\usepackage{hyperref}
\usepackage{amsmath,amssymb}
\hypersetup{
    colorlinks = true,
    urlcolor   = blue,
    citecolor  = black,
    linkcolor = black,
}
\usepackage{subcaption} 
\usepackage{caption}

\newcommand{\epz}{ak_0}
\newcommand{\vpz}{T_0}
\newcommand{\ap}{a_\mathrm{p}}
\newcommand{\epzp}{k_0\ap}
\newcommand{\bd}[1]{{\boldsymbol #1}}
\newcommand{\bau}{\bar{u}}
\newcommand{\baw}{\bar{w}}
\newcommand{\uL}{\bau_L}
\newcommand{\us}{u_s}
\newcommand{\usid}[1]{u_{s#1}}
\renewcommand{\k}{\kappa}         
\newcommand{\bu}{\bd{u}}
\newcommand{\bx}{\bd{x}}
\newcommand{\bk}{\bd{\k}}
\newcommand{\bom}{\bd{\omega}}
\newcommand{\bus}{\bu_s}
\newcommand{\babu}{\bar{\bu}}
\newcommand{\chbu}{\check{\bu}}
\newcommand{\tbu}{\tilde{\bu}}

\newcommand{\fac}{f_\mathrm{ac}}
\newcommand{\fenc}{f_\mathrm{enc}}

\newcommand{\chbom}{\check{\boldsymbol{\omega}}}
\newcommand{\chu}{\check{u}}

\newcommand{\rmi}{\mathrm{i}}
\newcommand{\rmd}{\mathrm{d}}
\newcommand{\rme}{\mathrm{e}}
\newcommand{\ouu}{\overline{u'u'}}
\newcommand{\oww}{\overline{w'w'}}
\newcommand{\ouw}{\overline{u'w'}}
\newcommand{\ouww}{\overline{u'w'w'}}
\newcommand{\ouwnum}{\overline{u'_1 u'_3}}

\newcommand{\ddt}[1]{\frac{\partial #1}{\partial t}}
\newcommand{\ddi}[1]{\frac{\partial #1}{\partial x_i}}
\newcommand{\ddj}[1]{\frac{\partial #1}{\partial x_j}}
\newcommand{\ddz}[1]{\frac{\rmd #1}{\rmd z}}

\newcommand{\rmddt}[1]{\frac{\rmd #1}{\rmd t}}
\newcommand{\half}{{\textstyle \frac12}}
\newcommand{\bomH}{\bom'^{(H)}}
\newcommand{\omH}{\omega'^{(H)}}

\newcommand{\homH}{\hat{\omega}^{(H)}}
\newcommand{\rms}{_\mathrm{rms}}
\newcommand{\urms}{u\rms}
\newcommand{\vrms}{v\rms}
\newcommand{\wrms}{w\rms}
\newcommand{\tu}{\tilde{u}}
\newcommand{\urf}{u_\mathrm{rf}}
\newcommand{\uwt}{u_\mathrm{wt}}
\newcommand{\buwt}{\bu_\mathrm{wt}}
\newcommand{\uwph}{\tilde{u}_\mathrm{ph}}
\newcommand{\LFOV}{L_\mathrm{FOV}}
\newcommand{\Tint}{T_\mathrm{int}}
\newcommand{\Tintgrp}{T_\mathrm{int,grp}}
\newcommand{\Tintreg}{T_\mathrm{int,reg}}

\renewcommand{\p}{\partial}
\newcommand{\betaf}{\beta_\mathrm{f}}
\newcommand{\betafz}{\betaf(0)}

\newcommand{\iA}{\textit{1.A}}
\newcommand{\iB}{\textit{1.B}}
\newcommand{\iC}{\textit{1.C}}
\newcommand{\iD}{\textit{1.D}}
\newcommand{\iE}{\textit{2.A}}
\newcommand{\iF}{\textit{2.B}}
\newcommand{\iG}{\textit{3.A}}
\newcommand{\iH}{\textit{3.B}}

\newcommand{\ib}{\textit{b}}
\newcommand{\ic}{\textit{c}}
\newcommand{\id}{\textit{d}}

\title{
Turbulence-induced anti-Stokes flow: experiments and theory}

\author{
Simen {\AA}. Ellingsen\aff{1}
  \corresp{\email{simen.a.ellingsen@ntnu.no}},
Olav R{\o}mcke\aff{1}, 
Benjamin K. Smeltzer\aff{1,2}, 
Miguel A.C.\ Teixeira\aff{3,4},
Ton S. van den Bremer\aff{5},
Kristoffer S. Moen\aff{1}, 
R. Jason Hearst\aff{1}
}

\affiliation{
\aff{1} Department of Energy and Process Engineering, Norwegian University of Science and Technology, 7491 Trondheim, Norway
\aff{2} Department of Ships and Ocean Structures, SINTEF Ocean, 7052 Trondheim, Norway
\aff{3} CEFT – Transport Phenomena Research Center, Faculty of Engineering, University of Porto, 4200-465 Porto, Portugal
\aff{4} ALiCE – Associate Laboratory in Chemical Engineering, Faculty of Engineering, University of Porto, 4200-465 Porto, Portugal
\aff{5} Department of Civil Engineering and Geosciences, TU Delft,
2628 CN Delft, The Netherlands
}

\begin{document}

\maketitle

\begin{abstract}
    We report experimental evidence of an Eulerian-mean flow, $\overline{u}(z)$, created by the interaction of surface waves and tailored ambient sub-surface turbulence, which partly cancels the Stokes drift, $u_s(z)$, and present supporting theory. Water-side turbulent velocity fields and Eulerian-mean flows were measured with particle image velocimetry before vs after the passage of a wave group, and with vs without the presence of regular waves. We compare different wavelengths, steepnesses and turbulent intensities. 
    In all cases, a significant change in the Eulerian-mean current is observed, strongly focused near the surface, where it opposes the Stokes drift. The observations support the picture that when waves encounter ambient sub-surface turbulence, the flow undergoes a transition during which Eulerian-mean momentum is redistributed vertically (without changing the depth-integrated mass transport) until a new equilibrium state is reached, wherein the near-surface ratio between $|\mathrm{d}\overline{u}/\mathrm{d}z|$ and $|\mathrm{d}u_s/\mathrm{d} z|$ approximately equals the ratio between the streamwise and vertical Reynolds normal stresses. This accords with a simple statistical theory derived here and holds regardless of the absolute turbulence level, whereas stronger turbulence means faster growth of the Eulerian-mean current. We present a model based on Rapid Distortion Theory which describes the generation of the Eulerian-mean flow as a consequence of the action of the Stokes drift on the background turbulence. Predictions are in qualitative, and reasonable quantitative, agreement with experiments on wave groups, where equilibrium has not yet been reached. Our results could have substantial consequences for predicting the transport of water-borne material in the oceans.
\end{abstract}


\section{Introduction}

The phenomenon of Stokes drift implies that periodic water waves in irrotational flow induce a net Lagrangian-mean transport along their direction of propagation \citep{vandenbremer17}. 
Although discovered theoretically a long time ago by \citet{Stokes1847}, Stokes drift has been elusive in laboratory experiments. A particular difficulty is in separating it from Eulerian-mean flows whose properties are determined by the boundary conditions of the flume itself \citep{Monismith2020}, although recent investigations in wave flumes where waves (or wave groups) propagate on initially quiescent water have provided convincing evidence \citep{grue17,vandenbremer19}.

Floating and suspended matter of small enough size, such as microplastics \citep{vanSebille2020}, oil spills \citep{Boufadel2021}, plankton \citep{Hernandez-Carrasco2018}, larvae and nutrients \citep{Rohrs2014}, are transported in the oceans with the Lagrangian-mean current $\uL$, equal to the Eulerian-mean current, 
$\babu(z)$,
plus the Stokes drift $\bus$, i.e., 
$ \babu_L(z) = \babu(z) + \bus(z)$. 
The importance of correctly modelling ocean transport makes prediction of the Lagrangian current a pivotal, but still open, question. For a monochromatic (nearly) linear wave of wavenumber $k_0$ and constant amplitude $a$ propagating in deep water the Stokes drift velocity along the direction of wave propagation is
\begin{equation}\label{eq:us}
    \us(z) = (\epz)^2c(k_0)\rme^{2k_0 z},
\end{equation}
where the intrinsic phase velocity (i.e., in the reference system where the surface is at rest) is $c$.
The phase velocity and group velocity, $c_g$, are given by
\begin{equation}\label{eq:disprel}
    c(k_0) = \sqrt{g/k_0} \text{ and } c_g(k_0) = \half\sqrt{g/k_0}
\end{equation}
where $g$ is the gravitational acceleration.
A contribution to oceanic transport velocities of magnitude $\us$ can be highly significant and must be taken into account when predicting, e.g., the development of oil spills or the fate of microplastic particles \citep{Hackett2006,Onink2019,Cunningham2022}. 

The na\"{i}ve approach for the modeller is to obtain the Lagrangian-mean current by adding the Stokes drift calculated from wave spectra
to the Eulerian flow, which is assumed to be unaffected by waves. However, several field studies have observed that waves appear to have no discernible effect on the Lagrangian-mean current, contrary to theory \citep{Smith2006,Lentz2008}. 
\citet{Smith2006} found that even short wave groups experience an Eulerian-mean current, which acts to entirely cancel the Stokes drift at the surface, and that the counter-current is strongly correlated with the presence of Stokes drift -- it appears only when the wave group is present and disappears once it has passed. 
Other observations find that the inclusion of Stokes drift does improve results, however, e.g.\ \citet{Rohrs2012}, who used drifters in coastal waters and employ a coupled wave--ocean model. 

Experiments of waves propagating on currents have also yielded results which are inconsistent with a simple addition of Stokes drift. In a careful laboratory study, \cite{Monismith07} found no change in Lagrangian-mean flow when waves were added, i.e., the Stokes drift is locally cancelled by an equal and opposite Eulerian flow. Moreover, reanalysis of previous experiments by \cite{Swan1990,Swan1990b}, \citet{Jiang1991}, and \citet{Thais1994} supported the same conclusion \citep[the observation seems to have been made also in the PhD work of][]{cowen96}. These results have remained something of a puzzle, as ``[n]o existing theory of wave--current interactions explains this behaviour'' as \citet{Monismith07} put it. Here, we demonstrate a mechanism that could resolve this conundrum at least in part. 

While being careful not to draw definite conclusions concerning the above-mentioned results, one might remark that some level of turbulence was likely to have been present alongside the waves in all these experiments.
The flow of \citet{Swan1990} passed through a honeycomb flow straightener which will have generated significant  turbulence levels, and the mean shear of the flow would provide further turbulence production, \citet{Jiang1991} discuss wave--turbulence interactions in their experiments in detail \citep[further detailed by][]{Cheung1988}, as do \citet{Thais1996} for the experiment of \citet{Thais1994}.
The experiment of \citet{Swan1990b} did not have imposed wind or current. Since measurements were made after the wavemaker had run for many hours, one may speculate that turbulence might well have been present, created, e.g., in the oscillating bottom boundary layer (the depth was less than a third of a wavelength) or thermal convection, with which the waves would have had ample time to interact, the flume being $18$\,m long. Naturally, this can be no more than speculation so long afterwards.

There have been indications that the interaction between waves and pre-existing turbulence will result in an alteration of the Eulerian current. Waves have the effect of reorienting and intensifying the turbulence beneath, as 
predicted theoretically \citep{Magnaudet1990,Teixeira2002}, and confirmed 
numerically \citep[e.g.][]{Guo2013,Tsai2017,Xuan2019,Xuan2020,Xuan2024}, experimentally \citep[e.g.][]{bliven1984,Cheung1988,Thais1996,Savelyev2012, Smeltzer2023} and in field studies \citep{Cavaleri1987,qiao2016}. 
Langmuir turbulence, the disordered pattern of long rolls approximately aligned with wind and waves due to Langmuir circulation formation at sea \citep{McWilliams1997}, was observed by 
\citet{Plueddemann1996}  to persist for up to a day after the wind had stopped, sustained by the surface waves the wind had created.

\cite{Pearson_2018} predicted with a theoretical argument that the interaction between waves and ambient turbulence would, on average, produce an Eulerian-mean current which opposes the Stokes drift near the surface. His paper has received little attention up until now, but in our theoretical work later, we shall draw heavily on his work and apply it to our own settings. Pearson's prediction shows that the Eulerian-mean current, $\bau(z)$, is opposite and similar to $u_s(z)$ near the surface but changes sign beneath the wave-influenced surface layer and integrates to zero as a function of depth. Although the wave--turbulence-induced Eulerian-mean current incurs no net change in mass transport, it will partly cancel the Stokes drift at the surface, and we therefore refer to it as an `anti-Stokes' current.

The picture which emerges is that, when waves and turbulence first meet, the combined flow goes through a transient `spin-up' period before a new 
quasi-equilibrium is reached, which includes the Eulerian anti-Stokes current. 
In our set-up, turbulence is created at the inlet of the mean flow and as it moves downstream towards the point of measurement it encounters waves and develops together with the velocity profile. When the time of effective interaction is short,  the result of partial spin-up is captured, while turbulence moving in the presence of waves for sufficiently long could reach a quasi-steady state before it is measured. The conditions for `short' and `long' interaction time are satisfied in our experiments, fo wave groups (Experiment 1) and regular waves (Experiments 2 and 3), respectively.

We review and extend statistical theory for both the transient and steady stages, and derive a theory based on Rapid Distortion Theory (RDT) which captures the underlying physics of the `spin-up' period and shows that the Eulerian acceleration of the current will exhibit approximately the same depth-dependent behaviour as the resulting current that we observe. The process is intimately related to the so-called `CL2' mechanism which creates Langmuir circulation \citep{Craik_Leibovich_1976}. It is worth noting that the CL2 mechanism, as lucidly reviewed by \cite{Leibovich1983}, requires the presence of an Eulerian-mean flow with slope (i.e., $\rmd\bau/\rmd z$) of the same sign as that of the Stokes drift profile, whereas the anti-Stokes current induced by wave--turbulence interaction, $\bau(z)$, has opposite slope. Thus, $(\rmd\bau/\rmd z)\cdot (\rmd\us/\rmd z) < 0$, which implies that the induced current tends to stabilise the combined system with respect to the CL2 instability.

\subsection{Possible previous observations}

Indications are that the same physical phenomenon has been at play in wave--current experiments performed in the context of studying the bottom boundary layer in shallow wave--current flows, motivated by understanding sediment transport (thus not highlighting the significance for ocean modelling). A string of independent measurements of the mean Eulerian flow in the presence and absence of waves by \citet{vanhoften77,bakker78,vandoorn81,kemp82,kemp83,Rashidi1992,klopman94,Mathisen1996} and \cite{Sing2017}, all showed that the waves caused a significant alteration of the mean flow near the surface, adding a contribution in the direction opposite to wave propagation in the near-surface region. More recent experiments also report the same \citep{Zhang2019,Peruzzi2021}. In addition to making the same observation, \citet{umeyama05,umeyama09a} found in his experiments that the vertical structure of the streamwise--vertical Reynolds shear stress depends strongly on the wave-propagation direction in the near-surface region.

Since these studies considered shallow currents where waves affect the bottom boundary layer, a direct comparison with our experiment in deep water is dubious, yet subsequent theoretical analysis gives reason to suspect a close connection. 
\cite{nielsen96} and \cite{dingemans96} propose two early explanations for the difference in Eulerian-mean current; the former relies on a force balance on average including the mean stress from waves and turbulence represented by eddy-viscosity, the latter on the creation of streamwise rolls due to the Craik--Leibovich vortex force due to the sidewall boundary layers, whose presence was observed by \cite{klopman94}. 
\cite{groeneweg98} developed a more sophisticated theory based on Generalised Lagrangian-Mean (GLM) theory, similar in spirit to the physical process we consider herein, but their analysis is not easy to compare with ours since it involves a set of coupled nonlinear differential equations and a complex turbulence model. Interestingly, \citet{groeneweg03} reconcile all three descriptions, at least qualitatively, by extending their GLM theory to three dimensions. \citet{Huang03} provided a careful theory, once more primarily concerned with the effect of waves on the bottom boundary layer, but also shedding light on the observed changes near the surface. They, too, use the simple eddy viscosity model of turbulence. In particular, they conclude that the mean wave-induced shear stress near the free surface is opposite in direction to the wave propagation, and is ``due largely to the distortion of eddy viscosity near the surface''. With a simple mixing-length model, \cite{umeyama05,umeyama09a} as well as \cite{Yang2006} find reasonable agreement with experiments. \cite{Olabarrieta2010} devise a simplified numerical model to avoid the restriction of low-steepness waves, and the perturbation theory of \cite{Tambroni2015} yields a model able to predict the Eulerian-mean velocity profile throughout the water column; both of the latter employ depth-dependent eddy-viscosity models. Crucially, all of these many model explanations depend on the simultaneous presence of waves and turbulence to explain the change in mean flow.

When reviewing previous numerical studies of waves in the presence of turbulence, one can also find evidence of the same Eulerian-mean current creation that we observe experimentally, even though the authors themselves have not discussed its significance especially. 
\citet{Borue1995} merely remark that the change in mean current near the surface should be studied further, while \citet{Kawamura2000} notes that the changes in current are higher the larger the Stokes drift magnitude, and discusses consequences for (Langmuir) turbulence production. Also \cite{Fujiwara2020a} find Eulerian-mean velocity profiles under Langmuir turbulence which are qualitatively very similar to those we measure under regular waves (their figures 6a and 9a) and report in section \ref{sec:Experiment2}, but do not discuss this point particularly.

\subsection{Other wave-driven currents}

Eulerian-mean flow driven by waves also occurs without pre-existing turbulence or vorticity from two separate mechanisms. First, to compensate for the divergence of Stokes drift on the group scale, Stokes drift in otherwise quiescent water must also be accompanied by an Eulerian return flow \citep[e.g.,][]{Longuet-higgins1962,vandenbremer16} for wave groups. In deep water, the depth-integrated Stokes drift and the depth-integrated Eulerian return flow are equal and opposite, 
and mass is preserved globally, not locally. This phenomenon is reviewed further in section \ref{sec:EulerianReturn}. 
The Stokes drift profile is highly concentrated near the surface whereas the return flow varies slowly with depth. The Eulerian-mean ``anti-Stokes'' current we observe varies rapidly with depth and cannot be explained by this mass conservation mechanism.
Second, there is also surface streaming driven by viscosity confined to
a thin viscous boundary layer beneath the surface, resulting from the imparting of wave momentum to the fluid as the waves decay \citep{longuethiggins53,Craik1982}. \citet{Tsai2017} and, recently, \citet{Fujiwara2024} studied how this current can also interact with waves to generate small-scale turbulence of Langmuir type via the CL2 mechanism. The viscous sub-surface layer is 
likely thinner than what our measurements can resolve, and, additionally, the resulting Eulerian-mean flow is directed along, not against, the direction of wave propagation. Third, the Earth's rotation causes a wave-induced Eulerian-mean flow that can exactly cancel the Stokes drift (for periodic waves and in the absence of viscosity), which is also known as the anti-Stokes flow \citep{Hasselmann_1970}. While this rotation-induced anti-Stokes flow could explain 
some field observations \citep{Lentz2008,Rohrs2012}, it cannot explain experimental results discussed above and presented herein
as our characteristic timescales are vastly smaller than Earth's period of rotation (the Rossby number is large).
Hence, neither of these three mechanisms can explain the observations just mentioned, nor those by, e.g., \citet{Monismith07} or indeed those we report herein.

\section{Experimental Methods} \label{sec:experimentalMethods}

\begin{figure}
  \centerline{\includegraphics[width=\textwidth]{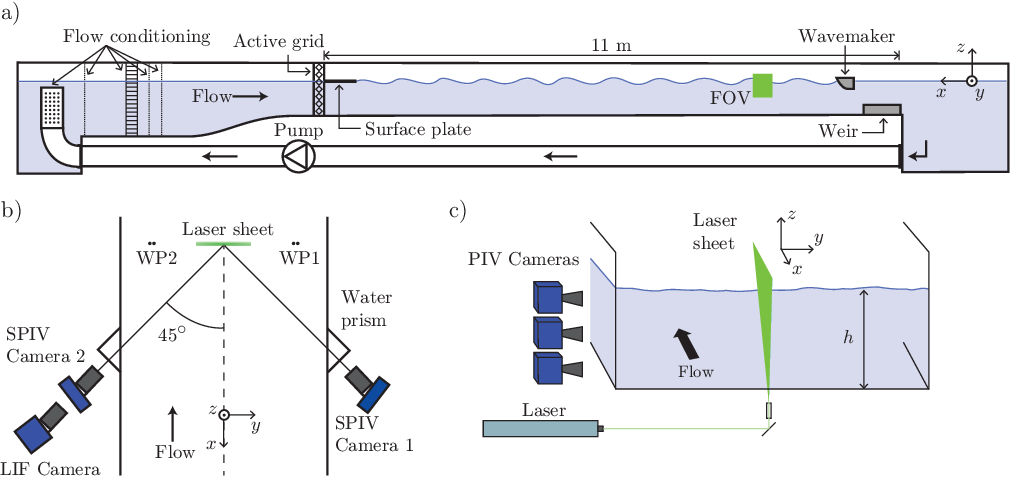}}
  \caption{Experimental set-up: (\textit{a}) side view of water channel 
  with flow from left to right and field of view (FOV) indicated with a green rectangle,
  (\textit{b}) top view of measurement region for the stereo particle image velocimetry (PIV) set-up in experiment 1, 
  including positions of wave probes (WPs) and laser-induced fluoresence (LIF) camera for surface detection; 
  (\textit{c}) longitudinal view of the planar PIV set-up in experiments 2 and 3. 
  Experiment 2 employed three stacked cameras as shown, whereas in
  experiment 3, a single PIV camera was used.
  }
\label{fig:exp}
\end{figure}

We report on measurements performed during three experimental campaigns in the water channel facility at 
NTNU Trondheim, shown in figure \ref{fig:exp}a. A pump system circulates water through the test section of dimensions $11.2$\,m $\times$  $1.8$\,m $\times$ $1.0$\,m (length $\times$ width $\times$ height). 

An active grid at the inlet of the test section allowed the turbulence to be generated and varied. 
The grid consists of square wings measuring $10$\,cm across the diagonal, attached to $18$ vertically and $10$ horizontally oriented bars, each controlled
independently by a stepper motor. 
Several different active-grid actuation cases were investigated, listed in Appendix \ref{app:grid}. The grid wings were rotated with random rotational velocity, acceleration, and period within set limits \citep{Smeltzer2023,Hearst2015}, 
or in one case flapped back and forth between two positions at irregular time intervals. 
The instantaneous rotation frequency of the grid wings varied about a mean active-grid frequency 
$\overline{f_G}$ by $\pm 0.5\overline{f_G}$ with a top-hat distribution. In experiments 1 and 2 (see sections \ref{sec:Experiment1} and \ref{sec:Experiment2}) a surface plate was mounted from the grid that extends downstream approximately $1$ m to dampen surface disturbances produced by the grid. 
A diagram of the set-up is shown in figure \ref{fig:exp}, and further details are given by \citet{Jooss2021}.

The bottom and sidewall boundary layers are thin enough not to reach the central part of the flow. They reach a maximum of approximately $20$\,cm at the point where velocity measurements were made, and are thinner upstream where waves and turbulence interact, hence essentially all turbulence which affects our experimental results has been generated by the grid and not boundary layers. The upstream distance within which waves and turbulence interact is denoted $\LFOV$ and is tabulated in Table \ref{tab:experiments}.

A plunger wavemaker at the downstream end of the test section ($10.2$ m from the active grid) was used to generate waves propagating upstream on the current.
Waves with a group velocity lower than the mean flow were unable to propagate upstream, thus preventing high-frequency wave noise and unwanted free harmonics (parasitic waves) from the wavemaker from entering the test section.
Wave properties were obtained from surface elevation measured by a pair of resistive wave probes (HR Wallingford) near the measurement position. 

As indicated by the coordinate system in figure \ref{fig:exp}a, the waves propagate in the positive $x$-direction, while the flow is in the negative $x$-direction. 
The mean free-surface level is at $z=0$.

\subsection{Experimental campaigns}

Three separate experimental campaigns were conducted between 2020 and 2023. Experiment 1 was also reported in \citet{Smeltzer2023}, where the focus was on the change in turbulent enstrophy and wave scattering.

\begin{table}
  \begin{center}
\def~{\hphantom{0}}
  \begin{tabular}{ccllccccccc}
      ID  & Cases & Type of waves & PIV  & PIV & $h$ &$f_0$ &$\fac$ &$N_\text{ens}$ & $T_\text{PIV}$& $\LFOV$ \\
          &       &               &type &plane & (cm)& (Hz) & (Hz)   &               & (s)           & (m) \\
      \hline 
      1 & \iA--\iD & Repeated groups& stereo & $yz$& 40& 1.02 & 8 &  $60^\dagger$& 10& 7.38\\
      2 & \iE, \iF & Regular waves& planar & $zx$& 80& 0.94, 1.16 & 0.86 & 1 & 2324 &8.5\\
      3 & \iG, \iH & Regular waves& planar & $zx$& 50 & 1.4 &15 & 32& 25 &8.5\\
  \end{tabular}
  \caption{
  Overview of the three experiments. Here, $h$ is the mean water depth, $f_0$ is the wave frequency in the laboratory frame (wavemaker frequency), $\fac$ is the PIV acquisition frequency, $N_\text{ens}$ is the number of repetitions per case, $T_\mathrm{PIV}$ is the duration of each acquisition interval and $\LFOV$ is the distance the flow travels as a free stream upstream of the measurement field of view.  
  $\dagger$: except case \iC.2 where $ N_\text{ens}=20$.
  }
  \label{tab:experiments}
  \end{center}
\end{table}

The three experiments investigate essentially the same phenomenon, but with fundamental differences in experimental design and the nature of the acquired measurement data. A summary is provided in table \ref{tab:experiments}.

\subsubsection{Experiment 1: Wave groups}
\label{sec:Experiment1}

Cases \iA-\iD\ in Tables
\ref{tab:experiments} -- \ref{tab:derived} 
are from Experiment 1. Wave groups were generated that propagated upstream atop the turbulent flows. The water depth $h$ was $0.4$\,m.
The velocity field was measured using stereoscopic particle image velocimetry (SPIV), measuring all three velocity components in a plane perpendicular to the mean flow located a distance $8.38$\,m downstream from the active grid (i.e., $83.8$ grid units), and $\LFOV=7.38$\,m downstream of the trailing edge of the surface plate. Two 25-megapixel cameras were mounted on either side of the test section, viewing the field of view at $\pm 45^\circ$ to the $x$-axis as shown in figure \ref{fig:exp}b). The field of view was $0.12\times0.14$\,m$^2$. Particle images recorded by the two cameras were processed using a final pass of $48\times48$ pixel interrogation window and a 50\% overlap, resulting in a velocity vector spacing of approximately $0.8$\,mm. The free surface intersection with the SPIV plane was detected from laser-induced fluorescence (LIF) images recorded by a camera viewing the plane at an oblique angle from the air side. A small amount of rhodamine-6G was added to the water generating image contrast between the air and water regions, and the free surface was detected from the image intensity gradient. Further details can be found in \citet{Smeltzer2023}.

\begin{figure}
  \begin{center}
      \includegraphics[width=\textwidth]{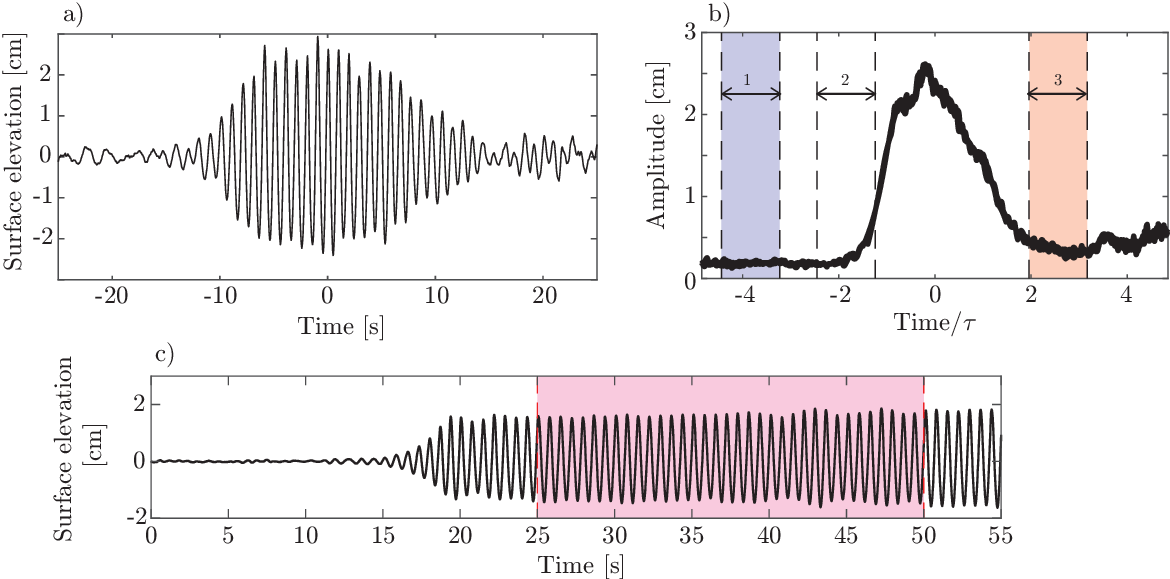}
  \end{center}
  \caption{(\textit{a}) Example surface elevation of a single wave group 
  from Experiment 1 as a function of time, measured by a wave probe at the measurement location. (\textit{b}) Example 
  from experiment 1  of an ensemble-average group surface elevation amplitude envelope as a function of time normalised by the 
  measured  group temporal width $\tau$ for case \iD (see \eqref{eq:wamp}). The time intervals for SPIV measurement (1-3) are shown 
  with vertical dashed lines.
  (\textit{c})   Surface elevation measurements of one ensemble from experiment 3 (cases \iG\ and \iH), which shows the onset of a regular wave train. The red box indicates the interval used for analysis. 
  }
\label{fig:int}
\end{figure}

For each wave group, SPIV/LIF measurements taken during three time intervals were used: 
well before the group arrived, and at the leading and trailing edges of the group envelope, referred to as intervals 1, 2, and 3, respectively, as shown in figure \ref{fig:int}b). 
We consider the difference in mean velocity between intervals 1 and 3 here, with interval 2 as a check to verify that the change is indeed due to waves. Values for all three intervals in all cases can be found in Supplementary Materials. 
The duration of the measurements for each interval, $T_{\mathrm{PIV}}$, was $10$\,s and PIV and LIF were sampled at $\fac=8$\,Hz.
After each group, residual waves from reflections were allowed to dissipate for approximately five minutes before the next wave group was generated. The above procedure was performed a total of $N_{\mathrm{ens}}=60$ times to produce ensemble statistics, except for the case \iC.2 which was performed $20$ times.  In case \iB, only vertically orientated bars of the active grid were actuated. For case \iA, the grid was stationary with the wings aligned with the flow in the position of least blockage (see also Appendix \ref{app:grid}).
The experimental conditions are listed in table \ref{tab:experiments}.

The wave groups were generated with the wavemaker motion having carrier frequency $f_0 = 1.02$\,Hz and a Gaussian amplitude envelope of the form:
\begin{equation}
    S(t) = S_0 \mathrm{exp}\left[-\frac{(t-T_\mathrm{wm}/2)^2}{2\tau_{\mathrm{wm}}^2}\right],
\end{equation}
for $0\leq t\leq T_\mathrm{wm}$, where $S_0$ is the peak stroke, $\tau_\mathrm{wm}=6$\,s is the group width in time 
as applied to the wavemaker signals and $T_\mathrm{wm} =24\,\mathrm{s}$ was the duration over which the wavemaker plungers were actuated. 
The surface elevation for one wave group measured at the SPIV measurement location is shown in figure \ref{fig:int}(a).

\subsubsection{Experiment 2: Regular waves}
\label{sec:Experiment2}

Cases \iE\ and \iF\ in tables \ref{tab:experiments} -- \ref{tab:derived} are from Experiment 2.
Regular waves were generated propagating upstream atop two different flows with comparable mean velocity but different levels of turbulence as controlled by the active grid. A planar PIV set-up with a light sheet, orientated in the streamwise--vertical ($xz$) plane, measured the in-plane streamwise and vertical velocity components. 
The field of view was centred $\LFOV=8.5$\,m downstream of the surface plate's trailing edge (see figure \ref{fig:exp}). 
Three cameras stacked vertically (from the top to bottom: two 16 megapixel cameras and one 5.5 megapixel camera) covered a field of view extending over the entire water depth ($h = 0.80$ m) of the channel, roughly $21.7$\,cm wide. 
The parameters for Experiment 2 are listed in table \ref{tab:experiments}.
For both flow cases, waves of two frequencies $0.94$ and $1.16$\,Hz each with three different steepness values were generated. For each case, $2000$ PIV images were acquired at a sampling frequency of $0.86$\,Hz; thus the total measurement time $T_{\mathrm{PIV}}$ was approximately $39$\,min. Using a final pass with $48 \times 48$ pixel interrogation window and a 50\% overlap resulted in a velocity vector spacing of approximately $1.6$~mm.  For both cases \iE\ and \iF, PIV measurements were also performed without wave generation to characterise ambient flow conditions, reported in Table~\ref{tab:mwt}.

\subsubsection{Experiment 3: Onset of regular waves}
\label{sec:Experiment3}

Cases \iG\ and \iH\ in tables \ref{tab:experiments} -- \ref{tab:derived} are from Experiment 3.
This set of measurements was taken at the highest acquisition frequency in our study ($15$\,Hz) during a $25$\,s interval
just after the first arrival of regular waves as indicated in figure \ref{fig:int}(\ic).
The water depth was $0.50$\,m, and a slower mean flow of $U_0=0.19$\,m/s was used compared with the other cases (see table \ref{tab:experiments}). The two cases are separated by the different sequences for the active grid. Case \iG\ is similar to case \iB\ as only vertically orientated grid bars were rotated at a random, top-hat-distributed rotation speed. Similarly for case \iH, only the vertical grid bars were actuated, made to flap over an angle of $\pm 60^\circ$ about the fully open position during random time intervals of a top-hat-distributed duration between $0.5$ and $1.0$ seconds. (A full list of experimental conditions is found in appendix \ref{app:grid}.) No surface plate downstream of the active grid was used in this experiment, and the field of view in the streamwise--vertical plane (see figure \ref{fig:exp}c) was centred a distance $\LFOV=8.5$\,m downstream of the grid.

The wavemaker was actuated at $f_0=1.4$ Hz.
Here, two steepnesses for each flow case are investigated, yielding four different steepness--turbulence combinations. For each of the four combinations, $32$ ensembles are measured, where the wave probe measurements of one ensemble are shown in figure \ref{fig:int}(c). 
In order to avoid wave breaking at the leading edge of the wave train, the wavemaker stroke is linearly ramped up to the set amplitude over a period of $6$ seconds. 

For the PIV measurements, a single 25-megapixel camera was used to cover the entire water column. Using a $64\times64$ pixel interrogation window and a 50\% overlap resulted in a final velocity vector spacing of approximately $3$~mm. The field of view is $0.45\, \mathrm{m}\times 0.5$\,m. PIV images were acquired at  $\fac=15$~Hz.
Also here, PIV measurements were performed without wave generation to characterise the ambient flow conditions, reported in table~\ref{tab:mwt}. For these background measurements $2000$ snapshots were captured at $1.0$\,Hz.


\section{Experimental measurements}

\subsection{Flow and wave characteristics}

The measured physical characteristics of mean flow, turbulence and waves are listed in table \ref{tab:mwt}, 
and some derived quantities we will make use of are quoted in table \ref{tab:derived} for convenience.

\begin{table}
  \begin{center}
\def~{\hphantom{0}}
  \begin{tabular}{cccccccc}
      Case   & $U_0$ & $k_0$ &$\tau$ & $[\urms, \vrms, \wrms]$ & $e$ & $L_x^x$ 
      &   Steepness      \\[3pt]
      &  $(\mathrm{m/s})$ &$(\mathrm{rad/m})$ & 
      $(\mathrm{s})$
      &$(\mathrm{cm/s})$ & 
      $(\mathrm{cm^2/s^2})$
      &$(\mathrm{cm})$ &   \\
      \hline 
      \iA & 0.34 & 9.5 &7.3& $[0.71, 0.68, 0.58]$  & 0.65 & $5.1$  &  $0.20$ \\
      \iB & 0.33 & 9.2 &7.2& $[1.1, 1.0, 0.73]$  & 1.3  & $26$ &  $0.20$ \\
      \iC & 0.33 & 8.9,9.0 &7.9& $[1.6, 1.2, 0.92]$  & 2.4  & $32$ &  $0.15,0.22$ \\
      \iD & 0.34 & 9.3 &7.8& $[1.7, 1.7, 1.3]$   & 3.7  & $20$  & $0.22$  \\
      \iE.1 & 0.30 & 6.1 &8.3& $[2.5,- ,1.8]$ & 6.4 &-   & $0.09, 0.14, 0.18$ \\
      \iE.2 & 0.30 & 12.1 &-& $[2.5,- ,1.8]$ & 6.4 &-  & $0.15,0.19,0.21$\\
      \iF.1 & 0.30 & 6.1 &-& $[1.6,-, 1.2]$ & 2.7 &-   & $0.07,0.12,0.17$\\
      \iF.2 & 0.30 & 12.0 &-& $[1.6,-, 1.2]$ & 2.7 &-  & $0.14,0.17,0.21$ \\
      \iG & 0.19 & 12.9 &-& $[1.2, -, 0.84]$  & 1.4 & 5.4 & $0.11, 0.18$ \\
      \iH & 0.19 & 13.0 &-& $[0.87, -, 0.70]$ & 0.87 & 5.1 & $0.11, 0.18$ 
  \end{tabular}  
  \caption{Measured flow quantities: mean current $U_0$, carrier-wave number $k_0$, measured group temporal width in the lab frame $\tau$, root-mean-square turbulent velocity components (subscript `rms'), turbulent kinetic energy $e$ (calculated for Experiments 2 and 3 as discussed in main text), and streamwise--streamwise integral scale $L_x^x$. Steepness is given as $\epzp$ for Experiment one where $\ap$ is peak amplitude, and $\epz$ in Experiments 2 and 3.
  Where several values of steepness are listed (cases \iC\ and \iE--\iH), these are referred to elsewhere as \iC.1, \iC.2, \iE.1.1, \iE.1.2, etc. 
  }
  \label{tab:mwt}
  \end{center}
\end{table}

\begin{table}
  \begin{center}
\def~{\hphantom{0}}
   \begin{tabular}{ccccccc}
      Case  &$\lambda_0$ & $c(k_0)$& $\us(0)$ &
      $\vpz$ & $\tau_0$& $|\urf|$\\[3pt]
      &$(\mathrm{m})$&  $(\mathrm{m/s})$ & $(\mathrm{cm/s})$& $(\mathrm{s})$& $(\mathrm{s})$ & $(\mathrm{mm/s})$ \\
      \hline      \iA &$0.68$ & $1.0$     & $4.1$         &$0.65$ &2.4 &  -\\
      \iB &$0.70$ & $1.0$     & $4.2$         &$0.66$ &2.6 &  - \\
      \iC &$0.71, 0.70$     & $1.0$& $2.3,5.1 $ &$0.67$ &2.9  & -\\
      \iD &$0.68$ & $1.0$     & $5.0$         &$0.66$ &2.8 &  - \\
      \iE.1 &$1.03$    & $1.3$     & $1.0, 2.5, 4.1$ &$0.81$&  - & $1.1, 2.5, 4.2$\\
      \iE.2 &$0.52$    & $0.90$    & $2.0, 3.3, 4.0$  &$0.58$&  -& $1.0, 1.7, 2.1$\\
      \iF.1 &$1.03$    & $1.3$     & $0.62, 1.8,3.7$ &$0.81$&  -& $0.6, 1.9, 3.8$\\
      \iF.2 &$0.52$    & $0.90$    & $1.8, 2.6, 4.0$  &$0.58$&  -& $0.9, 1.4, 2.1$\\
      \iG &$0.49$ & $0.87$    & 1.1, 2.8      &0.56&  - & $0.8, 2.2$\\
      \iH &$0.48$ & $0.87$    & 1.1, 2.8      &0.56&  - & $0.8, 2.2$\\
  \end{tabular}
  \caption{  Wave quantities derived from measured values in Table \ref{tab:mwt}. 
  Here $\lambda_0=2\pi/k_0$ is the carrier wavelength, $c(k_0)=\sqrt{g/k_0}$ is the carrier-wave phase velocity, $u_s(0)=(\epz)^2c(k_0)$ is the Stokes drift velocity of the carrier wave at the surface, $\vpz=\lambda_0/c(k_0)$ is the intrinsic wave period, $\tau_0$ the intrinsic group width from Eq.~\eqref{eq:intrinsicwidth}, and $\urf$ is the Eulerian return flow from Eq.~\eqref{eq:urf}.
  }
  \label{tab:derived}
  \end{center}
\end{table}

An assumption of deep water is well satisfied since in all our cases $k_0h\gtrsim 3.6$
hence $1>\sqrt{\tanh(k_0h)}> 0.999$.
Here and henceforth, $U_0$ is the mean absolute surface speed in the absence of waves 
(i.e., the mean surface velocity is $\mathbf{U}_0=(-U_0,0,0)$) and in all our cases the velocity profile is sufficiently depth-uniform that vertical shear affects wave dispersion negligibly \citep[e.g.,][]{Ellingsen2017}.
The wavenumber $k_0$ and the wavemaker carrier frequency $f_0$ are related approximately by $2\pi f_0 = \sqrt{gk_0} - U_0k_0$, or equivalently $f_0 = (c(k_0)-U_0)/\lambda_0$ with $c(k_0)$ from Eq.~\eqref{eq:disprel} and the wavelength is $\lambda_0=2\pi/k_0$. We will use the measured (as opposed to derived)
value of $\lambda_0$ and the wavenumber $k_0$ accordingly. Values for these quantities are listed in 
tables \ref{tab:mwt} and \ref{tab:derived}, respectively. 
The time it takes for the waves to propagate from wave-maker to the edge of the surface plate is 
around $40$\,s for Experiment 1, $25$\,s for Cases $\iE.1$ and $\iF.1$, $56$\,s for Cases $\iE.2$ and $\iF.2$ and $35$\,s for Experiment 3.

The root-mean-square (r.m.s.) of the turbulent velocity fluctuation after subtracting the average
is defined as $\bu\rms=[\urms,\vrms, \wrms]$ representing the streamwise, spanwise, and vertical components, respectively. The turbulent kinetic energy is as $e=\frac12 |\bu\rms|^2$. 
In cases \iE--\iH, only $\urms$ and $\wrms$ are available, so we use instead 
$e=\frac12 (\urms^2+2\wrms^2)$,
an assumption of cross-plane isotropy, as is the standard convention for grid turbulence \citep[e.g.,][]{comte-bellot1966,lavoie2005} and was observed for our facility by \citet{Jooss2021}. It is not quite satisfied for our flow as Table \ref{tab:mwt} shows for Experiment 1. However, given the inevitable anisotropy of the flow as the free surface is approached (also with no waves) and consequent depth variation of $e$ closer to the surface than approximately one integral scale \citep[the blockage layer, see, e.g.][]{aarnes2025}, representing turbulent kinetic energy (TKE) by a single number is always a simplified picture. The precise value of $e$ does not significantly affect our conclusions, and given that no simple expression for $e$ in terms of $\urms$ and $\wrms$ can reconcile all cases in Experiment 1, we choose to go with the conventional expression.

The mean velocity $U_0$ was calculated from averaging all streamwise velocities over the lower part of the field of view; averaging was performed over $-12$\,cm $<z<-5$\,cm for Experiment 1 and $-20$\,cm $<z<-10$\,cm for Experiments 2 and 3.
Full profiles of the Eulerian-mean velocity profiles for all experimental cases may be found in the Supplementary Materials.

When comparing turbulent quantities for the various cases one should bear in mind that these are measured at a fixed position in space, whereas the turbulence becomes gradually weaker as it travels downstream because of dissipation. The change in Eulerian-mean current is an integrated effect of wave--turbulence interactions that occurred upstream, where the turbulence intensity is in general a little higher. This is true of all cases, but because of the lower mean velocity, the turbulence that reaches the field of view in cases \iG\ and \iH\ has decayed for longer. 
A detailed study of turbulent decay in our lab with similar flow conditions as Experiment 1 was reported by \citet{Jooss2021}.
They found that for all cases, once the turbulence is fully developed TKE decays approximately as $\sim 1/x$ downstream for all cases. Thus, although the absolute values of TKE and Reynolds stresses are different at different positions, their relative magnitudes are expected to be constant. We consider only trends with respect to TKE, and ratios between Reynolds stresses which it seems reasonable to assume are well captured. 

In previous experiments by \cite{nepf1991} and \cite{klopman94,Klopman1997}, secondary motion in the form of a pair of streamwise rolls were measured when waves were propagated on a current, triggered by the CL2 mechanism. We could measure the cross-plane flow in Experiment 1, and while some weak in-plane mean flow was observed, of the order of $1$\,cm/s or less, the pattern was not consistent with a coherent pair of rolls. The resulting convection, in magnitude and direction, would be insufficient to convect boundary-layer turbulence into the near-surface regions where turbulence and waves interact to any noticeable degree. During the travel from grid to measurement position, the vortex force near the boundaries has far less time to accelerate a similar amount of water as in \cite{klopman94} and would likely not have time to develop. It is not perhaps so surprising that secondary flows would be different given that geometrical size and aspect ratios are rather different, and the current's travel time from grid to measurement point is less than half that in either of the mentioned flows even for our slowest flow in Experiment 3.

It was found by \citet{Smeltzer2023} that the turbulent integral length scale, characteristic of the largest turbulent structures,  is important for wave--turbulence interactions, especially so for the scatting of waves by turbulence. An advantage of our active-grid turbulence generation is that the integral scale can be varied independently of TKE. The most pertinent integral scale is related to correlation of streamwise turbulent velocity for points separated in the streamwise direction, $L_x^x$. Due to differences in the way the experimental data were acquired, no single method for estimating 
$L_x^x$ can be applied to all cases. Estimating the integral lengthscales from experimental data uniquely and quantitatively is notoriously difficult, and the various methods in common use produce quantitatively different results. Additional challenges pertain to active-grid turbulence due to the slow spatial decay of the autocorrelation function \citep{Puga2017, Mora2019}. In Experiment 1, the integral scale was estimated with a zero-crossing method as described by \citet{Mora2020}, and we used the same method to calculate the integral scale for Experiment 3, as listed in table \ref{tab:mwt}. 
Since the data in Experiment 2 have low time resolution the same method cannot be applied there, and using another, spatially based method would not give directly comparable numbers, so we provide no integral scale for Experiment 2. We note that integral scales are significantly shorter in Experiment 3 than in Experiment 1 at similar turbulence levels, which can be explained by the lower mean-flow velocity $U_0$.

Several practical aspects in evaluation of the wave and flow characteristics reported in table \ref{tab:mwt} differed for the three experiments, and are described in further detail below.

\subsubsection{Wave group measurements (Experiment 1)}

The mean flow and turbulence statistics without waves presented in Table \ref{tab:mwt} were evaluated over the first interval, where any influence from waves can be assumed to be negligible. The mean flow in the negative $x$-direction had approximately constant (absolute) value
$U_0$, found from averaging as described above. A small spanwise and vertical mean velocity was found, everywhere below $1$\,cm/s, mostly much less, which we attribute to the channel flow conditioning not being perfect (achieving a perfectly straight and uniform flow in a channel of our size is very challenging). Notably, unlike the streamwise velocity, the cross-plane velocities did not change after the passage of the wave group, which implies that there is negligible contribution secondary flow due to wave--current interactions near the side-walls mentioned above. 

The characteristic peak wave amplitude $\ap$ and the observed group temporal width $\tau$ were estimated from the ensemble-averaged amplitude envelope of the wave groups as measured by the probes near the SPIV/LIF laser sheet as seen in figure \ref{fig:int}(b). The average envelope was fitted to a Gaussian function of the form
\begin{equation}
    a(t) = \ap \exp\left[-\frac{(t-t_\mathrm{p})^2}{2\tau^2}\right],
    \label{eq:wamp}
\end{equation}
with $t_\mathrm{p}$ the temporal location of the group peak. The peak wave steepness is $\epzp$.

We define an intrinsic temporal group width $\tau_0$, listed in table \ref{tab:derived} as: 
\begin{equation}\label{eq:intrinsicwidth}
    \tau_0 = \tau\frac{c_g(k_0)-U_0}{c_g(k_0)},
\end{equation}
with $c_g$ defined in \eqref{eq:disprel}. The intrinsic temporal group width is expressed in a reference frame without mean flow and reflects the timescale during which the ambient turbulence interacts with the wave groups.

\subsubsection{Measurements with regular waves (Experiments 2 and 3)}

The ambient flow statistics were evaluated from PIV measurements acquired without waves. Similarly to Experiment 1, the mean flow profile varied only slightly across the measurement plane, and a representative absolute value $U_0$ is given in table \ref{tab:mwt}. 

The wave steepness $\epz$ was calculated using the average wave amplitude from the wave probe measurements in the proximity of the PIV measurement location. The amplitude of each individual wave oscillation varied slightly during the experiments, especially in cases with the highest level of turbulence, likely due to wave--turbulence interactions \citep[e.g.,][]{Smeltzer2023}. The variation is not of central interest to the present study, and thus only the mean steepness value is reported.  

The regular waves were present throughout the entire test section during the experiments. The grid-generated and measured turbulence thus interacted with the waves over a length $\LFOV$ as given in Sections \ref{sec:Experiment2} and \ref{sec:Experiment3}, with associated interaction time $T_\mathrm{int}=\LFOV/U_0$. 
The frequency at which turbulence encounters waves, i.e., as seen by the moving flow, equals the intrinsic wave frequency (in Hz)
\begin{equation}
    \fenc = f_0 + U_0/\lambda_0 = \sqrt{gk_0}/2\pi.
\end{equation}

\subsection{Measured change in Eulerian-mean velocity}

The measured changes in Eulerian-mean velocities presented below are of the order of millimetres per second, yet, while this is the same order of magnitude as our PIV measurement accuracy, one should bear in mind that the variance of an average from hundreds of independent measurements is far lower than that of single measurements. It is crucial that a careful analysis of errors and statistical convergence be performed in order to establish confidence that our results are accurate and reliable. In appendix \ref{app:errors}, we report results of these tests, supported by further data in the Supplementary Materials.

\subsubsection{Velocity change after the passage of wave groups (Experiment 1)} \label{sec:experiment_groups}

\begin{figure}
  \centerline{\includegraphics[width=.8\linewidth]{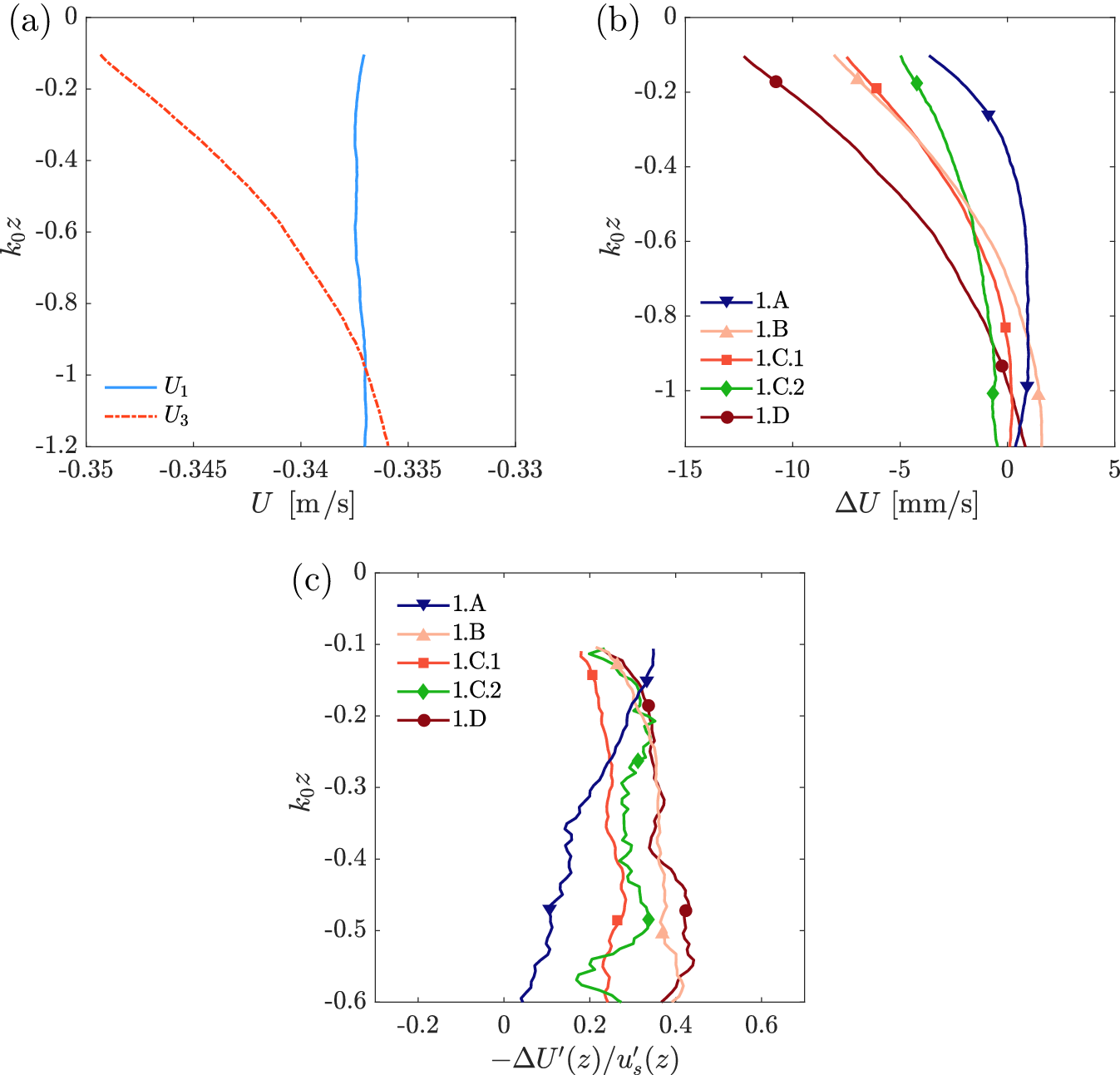}}
  \caption{
  Change in Eulerian-mean current due to the passage of a wave group. The waves travelled in the positive $x$-direction, against the current.
  (\textit{a}) An example of mean streamwise velocity depth profile 
  before the arrival of a wave group, $U_1(z)$, and after the group has passed, $U_3(z)$,  here for case \iD;  (\textit{b}) mean streamwise velocity difference $\Delta U = U_3-U_1$ as a function of depth for the flow cases of Experiment 1.   Error bars are omitted for visibility -- see analysis in Appendix \ref{app:errors}; (\textit{c}) the slope of $\Delta U(z)$ relative to the Stokes drift gradient (a prime denotes derivation with respect to $z$). Light smoothing (moving average with window size $8$\,mm) was applied to the curves in panel c for better visibility. 
  }
\label{fig:wavegroups}
\end{figure}

We now consider the measured Eulerian-mean velocity change for Experiment 1, i.e., cases \iA--\iD.
Vertically sheared flows in our laboratory have been found to be stable and therefore the wave-modified Eulerian-mean velocity profile remains close to unchanged throughout interval 3 \citep[also found to be true for strongly sheared currents, see Section 4 of][]{Pizzo23}.
In figure \ref{fig:wavegroups}(a) we show as an example the mean streamwise velocity profiles for case \iD\ during the three measurement intervals, denoted $U_I(z)$ for intervals $I = \{1,2,3\}$ (see figure \ref{fig:int}(b)), respectively. 
As can be seen, the mean flow speed had increased in the direction opposite to wave propagation after the passage of the wave groups. Similar plots for all cases are provided in the Supplementary Materials. Error bars are omitted from the figure for reasons of visibility, but discussed in Appendix \ref{app:errors}.
The net Eulerian-mean flow change $\Delta U=U_3 - U_1$ is shown in figure \ref{fig:wavegroups}(b) for flow cases \iA--\iD\ as expressed in the legend. For all cases, $\Delta U<0$ near the surface, and decays to small absolute values at depths $|k_0z|\gtrsim 1$. 

Interestingly, $\Delta U$ clearly depends on the level of ambient turbulence; while the waves have approximately equal properties in Cases \iA, \iB, \iC.1 and \iD, the velocity change is far higher at the highest turbulence level (case \iD) than at the lowest level (case \iA), with the intermediate cases \iB\ and \iC.1 in between. 
Moreover, comparing cases \iC.1 and \iC.2 where two wave steepnesses are tested on the same turbulent current indicates a positive correlation between $\epzp$ and current change $\Delta U$. Put together, the results in figure \ref{fig:wavegroups}(b) provide a strong indication that the change in current is due to an interaction between waves and turbulence.

We can exclude the possibility that the measured $\Delta U$ is due to the Eulerian return flow under groups of waves, as measured by \cite{vandenbremer19}. 
The return flow follows the group, i.e., it is very weak in intervals 2 and 3 which lie outside the main group itself. It is also near-uniform in depth, whereas the current measured in figure~\ref{fig:wavegroups}a is strongly depth dependent.

It is instructive to plot the ratio between $\rmd(\Delta U)/\rmd z$ and $-\rmd\us/\rmd z$ as a function of $z$, shown in figure~\ref{fig:wavegroups}(c), because it gives some indication of how far the turbulent flow has transitioned towards a new, quasi-equilibrium state. We will later present theory and evidence that at the end of the `spin-up' period, this ratio should, in the final state, be approximately equal to  $\urms^2/\wrms^2$ nearest the surface. Since our bulk turbulence is slightly anisotropic (in the cases reported in \citet{Jooss2021}, $1.2\lesssim\urms^2/\wrms^2\lesssim 1.4$), a final current change $|\Delta U|\gtrsim |\us|$ is expected nearest the surface. We shall later see that in the regular-wave cases where the equilibrium is likely reached, this relation holds well. Since none of the changes in currents in cases \iA--\iD\ are close to reaching these values, it seems that the passing of the wave group has not led to a wave--turbulence interaction of sufficient duration for a final state to be reached, and the flow is still relatively early in the `spin-up' stage. 

There are at least two striking observations to make in figure~\ref{fig:wavegroups}(c). First, with the exception of the low-turbulence case, \iA, the ratio between the slopes is close to constant with depth, which illustrates that $\Delta U \sim \exp(2k_0z)$ near the surface in these cases.
The scaling is not perfect, particularly at the shallowest depths; this should not be surprising since the depth dependence of wave--turbulence interaction should scale not only with the wavelength, but also the turbulent integral length scale, which delimits the vertical extent of the topmost layer where the kinematic boundary condition at the surface begins to limit the vertical extent of turbulent eddies \citep[the blocking effect, see, e.g.,][]{Teixeira2002}. 
Second, while the value of $\Delta U(z)$ after the passing of a group depends strongly on the turbulence level and steepness, there is no such trend for the relative slope ($-\Delta U^{\prime}(z)/u_s^{\prime}(z)$), Case \iA\ excepted. 

In Section \ref{sec:RDT} we will develop a RDT model describing the early onset of wave--turbulence interaction, which we can compare with the measurements in figure \ref{fig:wavegroups}b, given this evidence that the combined wave/turbulence flow is still far from fully developed in Experiment 1.

In conclusion, the evidence suggests that the change in Eulerian-mean current observed in our experiments is due to wave--turbulence interaction, and increases with increasing turbulence and increasing wave steepness.

\subsection{Velocity change under regular waves (Experiments 2 and 3)} \label{sec:Regular_waves_results}

Cases \iE--\iH\ all consider turbulence interacting with regular (i.e., continuous and periodic) waves.
For cases \iE.1--\iF.2, we evaluate the mean streamwise velocity in the presence of waves, and subtract off the mean velocity profile from the ambient flow case without waves. This velocity difference we define as $\Delta U(z)$. 
Although the flow is wavy, the time series is long enough for the Eulerian-mean current, averaged over all $2000$ PIV images, to be well converged in the sense that further measurements would affect it insignificantly. See details in Appendix \ref{app:errors}.

The turbulent current is affected by the waves during the time it takes it to traverse the test section, a distance $\LFOV$ as defined in Sections \ref{sec:Experiment2} and \ref{sec:Experiment3}. 
The interaction time is thus $T_\mathrm{int}=\LFOV/U_0$, 
approximately $29$\,s for Experiment 2 and $46$\,s for Experiment 3, the latter having a slower flow speed. This corresponds to a number of wave cycles, $T_\mathrm{int}/T_0,$ between $36$ (Cases \iE\ and \iG) and $83$
listed in table \ref{tab:RDTparams}. Due to the slower flow speed. Cases \iG\ and \iH\ interact for considerably longer than \iE\ and \iF, both in terms of absolute time (in seconds) and in terms of number of intrinsic wave periods 
$\vpz$, i.e., $\Tint/\vpz$ is the number of wave cycles that the turbulence has encountered.

\begin{figure}
    \begin{center}
    \includegraphics[width=\textwidth]{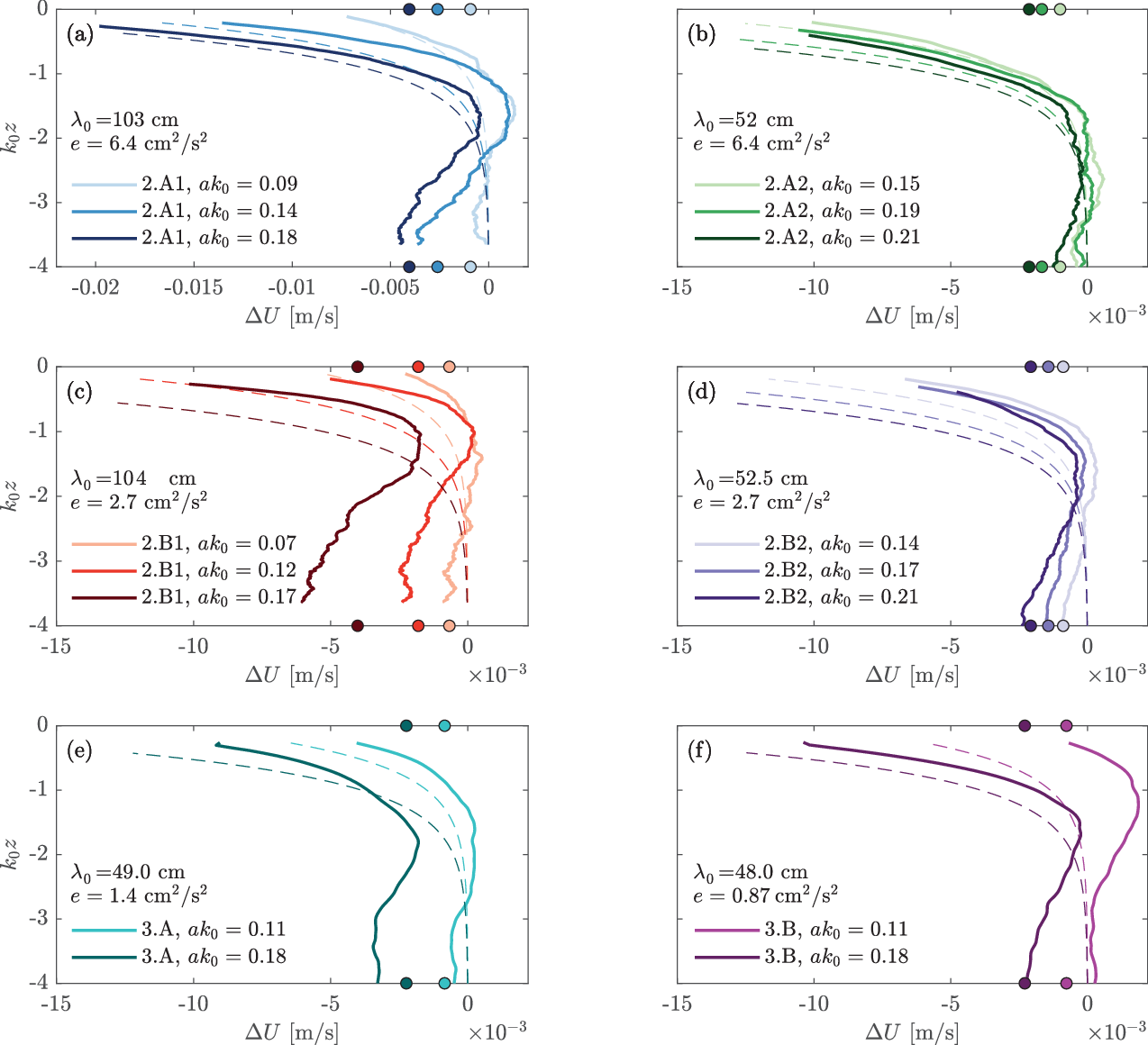} 
    \end{center}  
  \caption{
  The wave-induced current $\Delta U$ under regular waves as a function of depth $k_0z$ for cases \iE.1, \iE.2, \iF.1, \iF.2, \iG\ and \iH\ in panels (a)--(f), respectively. For each case, different wave steepness values   $\epz$  are shown as indicated in the legend. The dashed lines are the theoretical Stokes drift profiles at the same location for each case, shown as $-\us(z)$, that is, with opposite sign to the Stokes drift. The filled circles at $k_0z=-4$ and $0$ indicate the theoretical value of the Eulerian return flow, $\urf$.
  }
\label{fig:delU_lin}
\end{figure}

In figure \ref{fig:delU_lin} we show $\Delta U(z)$ for cases \iE--\iH, where the different wave steepness values $\epz$ are labelled in the legend. 
Several characteristic behaviours can be observed. First, all graphs have similar dependence on depth, with the highest absolute value nearest to the surface, decreasing down to a level where $k_0z$ is roughly in the range between $-1$ and $-2$, then turning and slowly becoming more negative again. (The non-monotonicity (turning) would not be visible in figure \ref{fig:wavegroups}, where measurements were only made for $k_0z>-1.2$). Indeed, some cases see the induced Eulerian-mean current take positive values over a certain depth range. \citep[One may note that the current profile is similar in shape and magnitude to those of][]{Rashidi1992}.
Second, there is again a clear tendency that higher steepness leads to a larger vertical variation of $\Delta U(z)$. 
Finally, there appears to be a near-constant offset between graphs (an addition to the mean flow), which increases with steepness, and we will find in Section \ref{sec:EulerianReturn} that it can be explained, at least in part, as the approximately depth-uniform Eulerian return flow $\urf$. 

The reverse of the theoretical Stokes drift $-\us(z)$ from  \eqref{eq:us} is plotted in all panels of figure \ref{fig:delU_lin} with dashed lines of the same colour for comparison, calculated for each case. For all cases we note that $|\Delta U|< \us$ nearest the surface, but that the depth variation $\rmd |\Delta U|/\rmd z\sim \rmd \us/ \rmd z$, is higher in some cases, smaller in others. We will return to these points later when comparing the results with theory.

\begin{figure}
  \centerline{\includegraphics[width=0.85\textwidth]{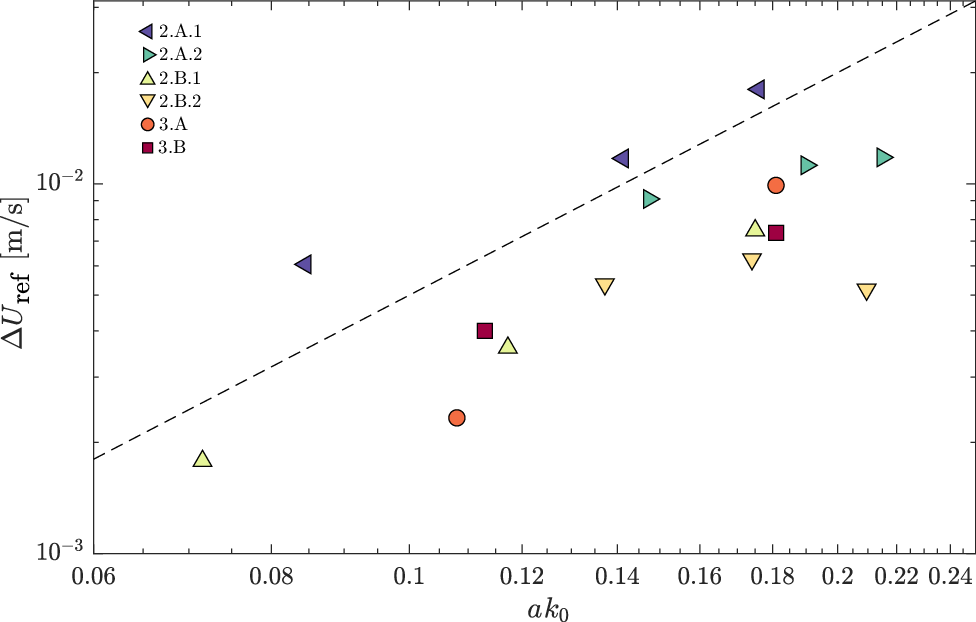}}
  \caption{The wave-induced current under regular waves at the `reference' depth $k_0z=-0.27$ as a function of wave steepness for cases \iE--\iH\,as indicated in the legend. The dashed line is proportional to $(\epz)^2$.
  }
\label{fig:delUmax}
\end{figure}

To further illustrate how the Eulerian-mean-current change depends on wave steepness, we plot the value of $\Delta U$ at a set reference depth as a function of steepness $\epz$ in figure \ref{fig:delUmax}. The value at the shallowest depth available for all cases is used, i.e. $k_0z=-0.27$. For cases \iE.1 and \iF.1, $\Delta U$ appears to scale as $(\epz)^2$, indicated by the dashed line. Cases \iE.2 and \iF.2 do not adhere to the scaling, possibly due to the high wave steepness values involved, so that interactions of order $(\epz)^3$ and higher may become significant. 
A further possible explanation is the lack of scale separation: the wavelength is likely only slightly larger than the integral length scale in these two cases (based on comparison with cases \iA--\iD\ in table \ref{tab:mwt}, which have similar $U_0$ -- unfortunately, a direct comparison of $L_x^x$ is not possible, as explained). The comparatively large integral length scale and high turbulence levels mean that stronger angular scattering of waves on turbulent velocity changes is to be expected \citep{VillasBoas20, Smeltzer2023}, a process which does not scale with steepness.
Figure \ref{fig:delUmax} should be interpreted only qualitatively, since the value of $\Delta U$ at a constant value of $k_0z$ is not entirely comparable between cases with different turbulence properties. As discussed in connection with figure \ref{fig:wavegroups}(c) and at length in Section \ref{sec:RDT}, $\Delta U$ depends not only on $k_0$ but also on the turbulent integral length scale and anisotropy, which varies between cases.

\section{Theoretical considerations}

The experimental results give reason to hypothesise that the changes in Eulerian-mean flow are a consequence of the encounter between waves and pre-existing turbulence. Before considering the turbulent current at all, however, we discuss the well-known Eulerian return flow which is present under wave groups also in quiescent water (Section \ref{sec:EulerianReturn}). We thereafter consider the model situation in which irrotational waves appear in the presence of pre-existing turbulence and the two begin to interact. In section \ref{sec:Pearson}, the argument made by \citet{Pearson_2018} is revisited and adjusted to our case, that the combined flow will undergo a transition---a `spin-up' as \cite{McWilliams1997} call it---to a new statistically steady state in which a new, depth-dependent Eulerian-mean current must be present. 
Next, a theory based on RDT is used to make more detailed predictions of the process during the `spin-up' period, predicting the depth profile of the `wave-Reynolds' stress which drives a mean current.

Throughout the theory sections, we assume that there is a negligible mean pressure gradient in the flow (apart from the local pressure gradient associated with a wave group) and that the mean flow after averaging over wave and turbulent motion (both time and ensemble average) lies in the streamwise--vertical plane, i.e.\ 
$\babu = (U_0,0,0)$, i.e., $\bar{v}=\bar{w}=0$, and all averaged quantities are presumed independent of the spanwise coordinate $y$. At the time and length scales we consider, the Coriolis force is irrelevant. 

We assume that a triple decomposition of the velocity field can be made according to 
\begin{equation}\label{eq:triple}
    \bu(\mathbf{x}, t)  = \babu(\mathbf{x}) + \tbu(\mathbf{x}, t) + \bu'(\mathbf{x}, t)
    \equiv \chbu(\mathbf{x},t) +     \tbu(\mathbf{x}, t), 
\end{equation}
where $\tbu$ and $\bu'$ are the contributions from waves and turbulence, respectively, and the Eulerian-mean flow $\babu$ changes slowly compared with the period and wavelength of an individual wave. The wave-filtered (Eulerian, turbulent) current is denoted $\chbu=\babu+\bu'$. 
Performing the decomposition in practice is non-trivial when the waves are not monochromatic and/or there is no clear scale separation in time or space. 
We tried and evaluated two different methods for separations---the widely used `phase-conditioned averaging' (PhCA) and proper orthogonal decomposition (POD)---and found the latter to perform better. A detailed comparison is found in Appendix \ref{app:POD}.

\subsection{Irrotational Eulerian return flow}\label{sec:EulerianReturn}

As is well known, the passage of a deep-water wave group does not entail a net mass flux, because the Stokes drift is cancelled in a depth-integrated sense by an Eulerian current in the opposite direction \citep{Longuet-higgins1962} known as the return flow. The current has been observed in laboratory studies, e.g. \citet{vandenbremer19}.

The return flow is found beneath a passing wave group, following the group and tending rapidly to zero ahead of and behind the group. For this reason, velocity measurements in cases \iA--\iD\, which are taken before and after the passage of the wave group, are expected to see negligible influence from such flow. Cases \iE--\iH, on the other hand, occur under regular waves -- in practice a very long wave group. Since the length of the ``group'' is far longer than the depth of the channel, the current will be approximately depth-uniform and, in the system following the mean current without waves, satisfying Lagrangian mass conservation, i.e., 
\begin{equation}\label{eq:urf}
  \urf =-\frac1h \int_{-h}^0 \us(z;h)\rmd z \approx -\frac1h \int_{-\infty}^0 \us(z)\rmd z = -
  \frac{k_0a^2}{2h}c(k_0),
\end{equation}
where we have assumed infinite water depth (i.e., $k_0h\gg 1$) for the Stokes drift profile, as we argued above, and inserted $\us$ from \eqref{eq:us}. The relation is valid when the depth is greater than about half a wavelength but much smaller than the group length ($\sim \LFOV$ in our experiment), both of which are well satisfied in our experiments. 
We note that \eqref{eq:urf} is only approximate, altered somewhat by the effect of the channel's bottom and sidewalls
(see e.g. \cite{vandenbremer17} for a discussion). 

\subsection{Statistical theory of Eulerian flow generation}\label{sec:Pearson}

\citet{Pearson_2018} argued that, when waves appear in a turbulent field, the combined flow will undergo a transition to a new statistically steady state. We will tailor his argument to our current application and discuss consequences in the context of our experimental observations. Momentum conservation, expressed through the time-averaged Navier--Stokes equations or the Craik--Leibovich equations as appropriate, implies that an Eulerian-mean current will manifest during the `spin-up' period, before a quasi-steady state is reached (in our case slowly decaying due to dissipation).

\subsubsection{Turbulent statistics in the transition period}\label{sec:Pearson-transient}

While quiescent-water irrotational waves and turbulence may each be steady in isolation (except for their decay due to dissipation), together they form a system in disequilibrium which will go through a transient change to a new quasi-steady state. 
This is similar to the model of `spin-up from rest' employed in theory for Langmuir turbulence \citep[e.g.,][]{McWilliams1997}. Revisiting and, to some extent, adapting the arguments of \citet{Pearson_2018} sheds light both on the early-stage `spin-up' stage and the final situation.
We first assume that a triple decomposition of the turbulent and wavy flow can be performed according to \eqref{eq:triple} (by no means a trivial point, an intricacy we shall return to later) and that all averaged quantities vary slowly as a function of $x$ over a wavelength so that their derivative can be neglected to leading order. Moreover, we assume that any changes in the mean flow develop slowly compared with a wave period.
This is reasonable because we have a strong indication that the interaction during passage of the wave groups in Experiment 1, which effectively lasts for approximately $\Tint\sim 7$ wave periods, is far from enough to reach the quasi-equilibrium state: the ratio $|\Delta U'(z)/u_s'(z)|$ only reaches approximately $0.5$ as figure \ref{fig:wavegroups}c shows, whereas we shall see in the next section (in turn backed by experiments) that in the final quasi-equilibrium state this fraction is $\gtrsim 1$. The wave field, and consequently the Stokes drift, changes slowly, so $\p_t\bus$ is negligible.
The flow is assumed to be unforced, without, e.g., wind stress, and there is no influence from buoyancy or the Coriolis force. The details of the dissipation are not important to the argument, so we shall not consider them beyond assuming that viscous decay is slow compared with the mean-flow effects of wave--turbulence interactions \citep[which is reasonable in light of the observations by][]{Jooss2021}, so that a quasi-steady state can be reached.  The final assumption is that $\babu$ and $\bus$ are both oriented along the $x$-axis and vary slowly except with coordinate $z$.

The turbulent flow is observed to develop slowly at the relevant scales, and can be assumed to be little affected by a Lagrangian average over a wave period. Wave-averaging the equations of motion yields the so-called Craik--Leibovich equation which under the above assumptions may be written in the form \citep[e.g.,][]{Suzuki2016}
\begin{equation}\label{eq:CL}
    \p_t \chbu +  (\chbu\cdot\nabla)\chbu = \bus\times\chbom -\nabla (\check{p} + \half \us^2 + \bus\cdot\chbu ) + \text{diffusion,}
\end{equation}
where $\chbom=\nabla \times \chbu =(0,\partial_z\bau,0) + \nabla\times\bu'$ denotes the (Eulerian) wave-averaged vorticity field. Taking the $x$-component and ignoring diffusion gives
\begin{equation}\label{CLx}
    \p_t \chu + \bau\p_xu'+ (\bu'\cdot\nabla)u'+ w'\p_z\bau \approx  -\p_x (p' + \us u' ),
\end{equation}
where we have used $\baw=0$. Performing a Reynolds average over turbulent motions (denoted with an overbar), while noting that $\nabla\cdot\bu'=0$ and that averaged quantities are independent of $x$, eventually yields
\begin{equation}\label{eq:RANS2}
    \p_t \bau \approx - \partial_z \ouw.
\end{equation}

Equation \eqref{eq:RANS2} has two important consequences. First, that an inhomogeneous field of turbulence will drive a mean current for as long as a vertical gradient of the shear stress exists (and dominates over viscous forces). Second, when a steady state has been reached, i.e., $\partial_t\bau=0$, then $\ouw$ is approximately constant with respect to depth \citep[see also][for more discussion]{Pearson_2018}, or, more precisely, its vertical gradient is balanced by viscous diffusion. Physically, $\ouw\neq 0$ represents a vertical redistribution of streamwise momentum, allowing $\bu=\bu(z)$ to transition from its initial value to a presumed final steady-state value, so when the transition period is ended, this shear stress should thus be small compared with TKE.
Equation \eqref{eq:RANS2} paves the way for further analysis of the `spin-up' of the Eulerian-mean current, because the right-hand side can be related to the underlying physical process using a model based on RDT.

\subsection{Statistics in the quasi-equilibrium state}\label{sec:Pearson-steady}

Following \citet{Pearson_2018} and \citet{Harcourt2013}, it can be readily argued that the development of a mean flow via the turbulent shear stress $\ouw$ is due to interaction between pre-existing turbulence and Stokes drift.  

Multiplying the $x$-component of \eqref{eq:CL} by $w'$ and the $z$-component by $u'$, adding them together and averaging, yields
\begin{equation}\label{eq:uwtransp}
    \ddt{\ouw}+\oww \ddz{\bau} =- \ouu\ddz{u_s} - \ddz{}\ouww + \mathrm{diffusion},
\end{equation}
where we have employed the chain rule and the fact that $\nabla\cdot\bu'=0$. The diffusion term contains viscous dissipation and turbulent pressure fluctuations, both of which may reasonably be assumed to be small in an oceanographic setting \citep{Harcourt2015, Pearson_2018} 
\citep[unfortunately we cannot directly ascertain how accurate this assumption is in the present context; see also][]{Pearson2019}. 
We should bear in mind that this assumption is questionable for our experiment, where waves are shorter and turbulence levels considerably higher than in typical field settings. This is especially true for cases \iE.2 and \iF.2. For instance, angular diffusion of the waves can be highly significant when the waves are not long compared with the turbulent integral scale as found by \citet{Smeltzer2023} which would cause a spatially variable Stokes drift. 
The approximation is more justified the larger the Stokes drift ($\sim (\epz)^2c(k_0)$) is compared with turbulent velocities ($\sim \urms$). In cases \iE.2 and \iF.2 the turbulence is strong and the waves short, and turbulent diffusion and pressure correlation terms may not be negligible compared with the Stokes drift contributions even at surface level. Unfortunately, we are not in a position to quantify the diffusive terms in \eqref{eq:uwtransp} from our data.

We now assume that a quasi-steady state has been reached, so that the explicit time derivative of $\ouw$ is negligible. The term $\ouww$ represents net vertical transport, and is expected to become very small in a quasi-steady state. Assuming the diffusion is small, it follows that
\begin{equation}\label{eq:EulerStokes}
    \oww \ddz{\bau} \approx - \ouu\ddz{\us},
\end{equation}
which is to say that an Eulerian-mean current must have been created with the opposite sign compared with Stokes drift. 

Note carefully that the relation \eqref{eq:EulerStokes} will only hold as long as the right-hand side term is large enough to dominate over the terms in \eqref{eq:uwtransp} we have neglected. Since the Stokes drift decreases exponentially with depth, our assumptions will surely be highly suspect deeper than $k_0z \sim -2$, possibly above if the turbulence is strong.

In a situation with two horizontal directions (such as in an ocean wave model), the above argument easily generalises to 
\begin{equation}\label{eq:EulerStokes2}
    \oww \ddz{\babu_h} \approx - \overline{\babu_h'\cdot\babu_h'}\ddz{\bus}
\end{equation}
with $\bus$ and $\babu_h$ lying in the horizontal plane. 
We emphasise that if a constant, depth-uniform current is initially present (as in our experiment), the current $\babu$ in \eqref{eq:EulerStokes} and \eqref{eq:EulerStokes2} is the \emph{change} in current due to wave--current interaction, or alternatively the current measured in the reference system following the original Eulerian-mean flow.

Beneath a free surface, turbulence is not isotropic, yet $\ouu$ and $\oww$ can be expected to be of the same order of magnitude 
except very near the mean surface level where $\oww$ tends to zero. 
It is worth noticing that \eqref{eq:EulerStokes} is not a perfect cancellation between Eulerian flow and Stokes drift as suggested by the experimental (re)analysis of \cite{Monismith07}, but when turbulence is close to isotropic, the remaining Lagrangian-mean current could be difficult to distinguish from zero in an experiment. On the basis of available evidence it seems a reasonable conjecture that the wave--turbulence-generated anti-Stokes flow explains, at least partly, this surprising cancellation. 

A key aspect to notice in \eqref{eq:EulerStokes} for modelling purposes is that the degree of cancellation of the mean Lagrangian current does not depend on the overall turbulence level, only on the anisotropy of turbulent fluctuations within the near-surface layer of thickness $\sim 1/k_0$ where Stokes drift is non-negligible. On the other hand, the rate of growth of the Eulerian current is proportional to $\partial_z\ouw$ and is higher the more intense the pre-existing turbulence is. In an ocean setting some level of turbulence is nearly always present, so the partial cancellation of Stokes drift according to \eqref{eq:EulerStokes} is to be expected.

\subsubsection{Comparison with experiment}\label{sec:ExpThComparison}

\begin{figure}
    \begin{center}
     \includegraphics[width=\textwidth]{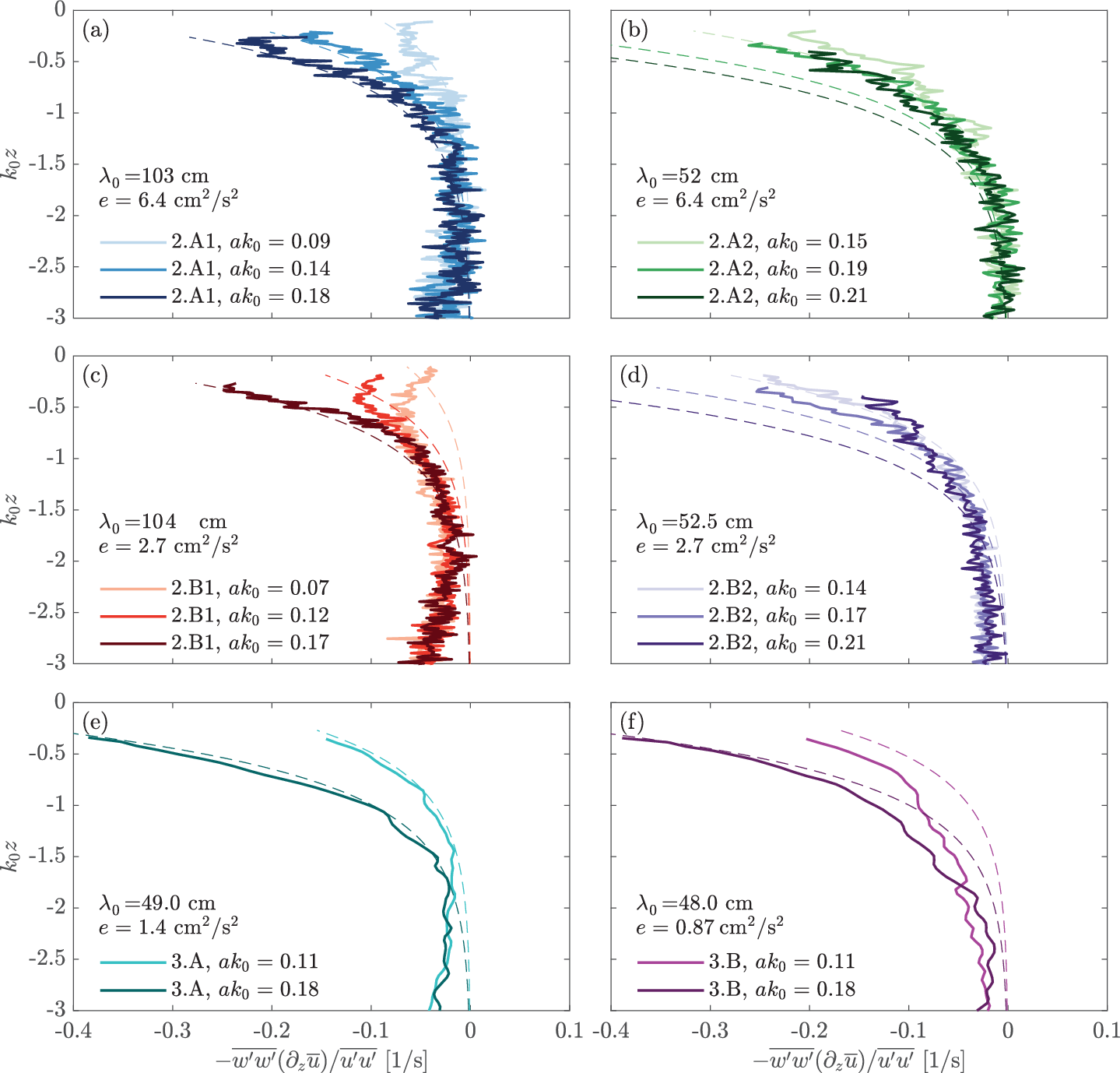}
    \end{center}
    \caption{ Test of equation \protect\eqref{eq:EulerStokes} for cases \iE\ to \iH. 
    The theoretical Stokes drift gradient $\rmd \us/\rmd z$ (the derivative of equation \ref{eq:us}) is shown as dashed lines of corresponding colour.
    Separation of turbulence from waves was performed with POD, discussed in Appendix \ref{app:POD}. 
    }
    \label{fig:StokesDriftVsEuler_EH}
\end{figure}

In order to test \eqref{eq:EulerStokes}, it is necessary to extract the variances $\ouu$ and $\oww$ from the wave--turbulence PIV data. 
Accomplishing this with the best possible accuracy is paramount in order to estimate turbulent second-order moments, because the wave velocities far exceed those of the turbulence near the surface, so even a small percentage of the wave motion erroneously identified as turbulence could lead to considerable overestimates of $\ouu$ and $\ouw$ (less so $\ouw$ due to the phase difference between $\tu$ and $\tilde{w}$). 

In cases \iG\ and \iH\ one could employ the now-standard technique of phase-conditioned averaging (PhCA) \citep{umeyama05,Buckley2017} to average out the wave motion, because the wave phase at each point in time and space can be estimated. For cases \iE\ and \iF\ the task is more difficult because measurements are made with frequency lower than the wave's, and no phase information is available (Experiment 2, when performed, was not planned with this task in mind). 
Estimates of $\ouu$ and $\oww$ can be made in all cases regardless of acquisition rate using the method of Proper Orthogonal Decomposition (POD) which decomposes the measured velocity field into spatial modes ordered according to their `energy' content. A more detailed discussion of the performance of this method and validation may be found in Appendix \ref{app:POD}. In cases \iG\ and \iH\ direct comparison with PhCA can be made; for the purpose of comparing with \eqref{eq:EulerStokes}, the two methods give very similar, but not quite identical results.

Although PhCA is in common use while POD has not been employed for this purpose previously to our knowledge, there are fundamental reasons why a data-driven method such as POD,
in general, is more appropriate. In particular, PhCA relies on assumptions which typically, including in our case, are not fully met.
A further discussion of this point is found in Appendix \ref{app:POD}, where we argue that POD is the preferred method of decomposition also in cases \iG\ and \iH. 

The two sides of \eqref{eq:EulerStokes} are compared in figure~\ref{fig:StokesDriftVsEuler_EH} for cases \iE\ to \iH, the cases whose measurements were taken in the streamwise--vertical plane so that waves and turbulence can be separated. The POD procedure was employed for triple decomposition to evaluate turbulent second moments. 
The dashed lines show the vertical derivative of the Stokes drift, $\rmd u_s/\rmd z=2\sqrt{gk_0}(\epz)^2\exp(2k_0z)$, while the solid line is the measured value of $-(\oww/\ouu)\rmd\bau/\rmd z$, which is hypothesised 
to equal the Stokes drift gradient in the quasi-steady state. 
For completeness, the Reynolds stresses themselves are plotted in Appendix \ref{app:Reynolds}.

Panels (e) and (f) of figure \ref{fig:StokesDriftVsEuler_EH} are from Experiment 3 where the separation of waves and turbulence could be performed and validated with confidence, as discussed in appendix \ref{app:POD}, but for cases \iE\ and \iF\ in panels (a)--(d) the uncertainly is difficult to quantify. For this reason, considerable caution should be exercised when interpreting the results in figure~\ref{fig:StokesDriftVsEuler_EH}(a)-(d), particularly nearest the surface where the wave motion is most energetic, so that even a small percentage of wave motion remaining in the calculation of $\ouu, \oww$ could cause a significant overestimation. We have decided to include the figures nevertheless, to illustrate the overall similarity of trend. 
 
Given the highly irregular and unsteady nature of our strongly turbulent flow, the agreement is striking in many of the cases. 
As argued above, the assumptions behind \eqref{eq:EulerStokes} are expected to be reasonable when Stokes drift is sufficiently high compared with turbulent velocities, i.e. $k_0z \gtrsim -2$ and sufficiently large $\epz$, and subject to corrections if diffusion levels are high. The relatively poor fit for cases \iE.2 and \iF.2 is therefore not very surprising for the latter reason. The very low values of $\epz$ in cases \iE.1.1 and \iE.2.1 is a likely reason why \eqref{eq:EulerStokes} is unlikely to be reasonable for these cases -- again the Stokes drift term in \eqref{eq:uwhom} may not dominate over terms which are neglected. Indeed, the good performance in case \iG.1 is more striking, but tallies with the fact that TKE is only half that of case \iF\ and less than a quarter of that of case \iE, with correspondingly less diffusion. 

Some further words of caution are warranted. As discussed previously, turbulence was measured only at one position near the downstream end of the channel, while the wave--current interaction occurred upstream. We therefore cannot guarantee that the ratio $\ouu/\oww$ has been the same throughout the time of interaction -- indeed the theory of \citet{Teixeira2002} indicates this ratio has increased in magnitude near the surface during the spin-up period due to wave--current interaction. 
We hesitate to offer an explanation for the surprising fact that the experimental curves in panel figure~\ref{fig:StokesDriftVsEuler_EH} appear to curve back near the surface, corresponding to a similar behaviour of the fraction $\ouu/\oww$ (see also figure \ref{fig:Appendix_ReynoldsStress}(i)), yet we remind the reader of the difficulty of separating waves and turbulence near the surface for these cases; we cannot rule out that the effect is spurious.

\subsection{A model based on RDT}\label{sec:RDT}

In section \ref{sec:experiment_groups}, we measured the change in Eulerian current after the passage of a wave group. Unlike for the regular wave cases, the shear of the resulting current is considerably smaller than the gradient of the Stokes drift, so the two sides of \eqref{eq:EulerStokes} are highly dissimilar, which signifies that the influence of the group of waves has not brought the flow near the quasi-steady state. Since evidence suggests that the flow is still `spinning up' when the group has passed, a theory which describes the development soon after the onset of waves is expected to describe the physical process in further detail.

The early onset of the (change in the) Eulerian current is governed approximately by \eqref{eq:RANS2}, and we will use RDT to study how its right-hand side---that is, the Reynolds stress $\ouw$---depends on depth $z$ and time while changes are still moderate. The theoretical predictions will be compared with the observations made for cases \iA--\iD, shown in figure~\ref{fig:wavegroups}(b). 
\citet{Teixeira2002} showed using RDT that a shear stress that varies in the vertical is generated by the passage of a progressive monochromatic surface wave over isotropic turbulence. We will show next that something similar occurs for a finite wave group, such as considered in the experiments of the present study.
Since the effect we consider is inviscid in nature, we will neglect viscosity in the following.

The RDT is a theory first proposed by \cite{Batchelor1954} where the straining of turbulence due to the distortions from the surrounding flow is assumed to dominate over that due to turbulence acting on itself, so that the term $(\bu'\cdot\nabla)\bu'$ is negligible in the Navier--Stokes equation, yielding a linearised theory. 
The leading cause of change for the turbulence is presumed to be the mean Lagrangian motion of turbulence-containing parcels of fluid due to the larger-scale surrounding motion. 
The RDT approach is formally valid whenever the distortions applied to turbulence are sudden. This amounts to assuming that the distortions (for example, mean-flow gradients) are applied over a time shorter than an eddy turn-over time. The spectral formulation of RDT also requires that there is a spatial scale separation between the turbulence and the mean flow. In the present wave-associated mean flow, this corresponds to $\lambda \gg L$ where $\lambda$ is the wavelength of the waves and $L$ is the integral length scale of the turbulence (cf. \citet{Teixeira2002}). 
Both of these criteria are reasonably well satisfied in the experiments \iA--\iD\ where the relevant turbulent lengthscale, $L_x^x$ is at most half a wavelength. Visual inspection of the velocity field clearly shows that the typical coherent eddies are much smaller than $\lambda_0$. The values of $L_x^x$ and details of its calculation with the method of \citet{Mora2020} were reported in \citet{Smeltzer2023}, and are quoted in Table \ref{tab:mwt}.
In practice, RDT is known to provide useful results even when these conditions are not strictly fulfilled \citep[e.g.][]{Hunt1990,Mann1994,Cambon1999}.

In the present case, the effect that is essential in order to explain the generation of a Eulerian-mean current is not related to the individual wave oscillations but rather to the systematic tilting and stretching of the vorticity of the turbulence by the Stokes drift of the wave. Therefore, we adopt a linearised version of the Craik--Leibovich equation, \eqref{eq:CL} whose key term is the `vortex force' $\bus\times\chbom$.
Applying a simplified phase-averaged formalism is a good approximation because individual wave cycles have little effect on the net stretching and tilting of vortices, and it is the integrated effect of the whole wave group which matters; this is illustrated, e.g., in figure 3 of \cite{Teixeira2010} (regular waves are considered there, but the arguments hold equally for the relatively long wave groups we consider here). Only a small uncertainty is thereby introduced which is far smaller than the quantitative effects of the simplifications underlying RDT itself. 
The corresponding vorticity equation, which will be used in RDT, is
\begin{equation}
    \ddt{\bom'} + (\bus\cdot\nabla)\bom' = (\bom'\cdot\nabla)\bus.
\label{vorteq}
\end{equation}
In (\ref{vorteq}), $\bom' = \nabla\times\bu'$ is the turbulent vorticity. As before, $\bus$ is orientated along the $x$ axis and is independent of $x,y$ and $t$. 
The mean background current is assumed to be initially depth uniform; hence, it has only trivial effect on the flow system, and we set it to zero by a change of coordinate system.
It should be noted that these equations represent the effect of the Stokes drift of an irrotational surface wave; the rotational correction to the wave motion due to the small change in Eulerian current \citep[see][]{Ellingsen2016} is neglected.

Let us first consider the turbulence away from the air-water interface, which to a first approximation can be considered homogeneous and isotropic, being denoted by the superscript $(H)$. 
In this section, it is useful to let the subscript $i=1,2,3$ denote the component of a vector or tensor along directions $x,y$ and $z$, respectively. 
The turbulent, homogeneous (or bulk) vorticity 
$\bomH$ is expressed as a three-dimensional Fourier integral, as
\begin{equation}
    \omH_i(\bx,t)=
    \iiint \homH_i(\bk,t) \, \rme^{\rmi \bk \cdot \bx} \, \rmd \k_1 \, \rmd \k_2 \, \rmd \k_3,
    \label{fourint}
\end{equation}
where the hat denotes a Fourier transformed turbulent perturbation quantity and  
$\bk=(\k_1,\k_2,\k_3)$ is the wavenumber vector.

Two equations result from (\ref{vorteq}) together with (\ref{fourint}):
\begin{align}
    \rmddt{\homH_i}&= 
    \ddj{u_{si}}
    \homH_j ,     
    \label{spectral1} \\
    \rmddt{\k_i} &= - \ddi{\usid{j}}\k_j.
    \label{spectral2}
\end{align}
Since $u_{si} = \us \delta_{1i}$, (\ref{spectral1})-(\ref{spectral2}) reduce to
\begin{subequations}
\begin{align}
    &\rmddt{\homH_1} = \ddz{\us}\homH_3,   
    \quad     \rmddt{\homH_2}=0,\quad \rmddt{\homH_3}=0,
    \label{vorteqft} \\
    &\rmddt{\k_1}=0, \quad \rmddt{\k_2}=0, \quad \rmddt{\k_3}=-\ddz{\us}\k_1.
    \label{wavenumeq}
\end{align}
\end{subequations}
These equations can be integrated in time to yield
\begin{subequations}
\begin{align}
    &\hat{\omega}_1^{(H)}(t)=\hat{\omega}_{10}^{(H)}+\hat{\omega}_{30}^{(H)} \int_0^t \ddz{\us}\rmd t',
    \quad \hat{\omega}_2^{(H)}(t)=\hat{\omega}_{20}^{(H)}, \quad \hat{\omega}_3^{(H)}(t)=\hat{\omega}_{30}^{(H)}, \label{sols1} \\
    &\k_1(t)=\k_{10}, \quad \k_2(t)=\k_{20}, \quad \k_3(t)=\k_{30}-\k_{10} \int_0^t \ddz{\us}\rmd t', \label{sols2}
\end{align}
\end{subequations}
where the subscript `$0$' applied to a variable denotes its value at the initial time, before the turbulence is distorted. It is convenient to define
\begin{equation}
\beta = \int_0^t \ddz{\us}\rmd t'  = \ddz{\Delta x},
\label{eq:stokesdisp}
\end{equation}
where $\Delta x(z,t)$ is 
the total fluid parcel displacement in the $x$-direction associated with the Stokes drift.

In order to calculate statistics of the turbulent velocity, it is necessary to relate the turbulent velocity fluctuations before and after distortion. 
Continuity of $\bu'$ implies 
\begin{equation}
    \nabla^2 \boldsymbol{u}' = - \boldsymbol{\nabla} \times \boldsymbol{\omega}' \label{poisson}
\end{equation}
which in spectral space becomes
\begin{equation}
\hat{u}_i^{(H)} = {\rm i} \varepsilon_{ijl} \frac{\k_j}{\k^2} \hat{\omega}_l^{(H)}, \label{poissonft}
\end{equation}
where $\varepsilon_{ijl}$ is the Levi-Civita permutation symbol and $\k = |\bk|$.

We wish to relate the turbulent velocity after distortion by the Stokes drift to the velocity of the initial undistorted turbulence. 
The Fourier transform of the velocity of the homogeneous turbulence after distortion by the Stokes drift can be related to the corresponding Fourier transform of the vorticity using (\ref{poissonft}). In turn, the vorticity of the homogeneous turbulence after distortion can be related to the initial turbulent vorticity using (\ref{sols1})-(\ref{sols2}). Finally, the initial turbulent vorticity can be related to the initial turbulent velocity through
\begin{equation}
    \hat{\omega}_{i0}^{(H)} = {\rm i} \varepsilon_{ijl} \k_{j0} \hat{u}_{l0}^{(H)},
    \label{vortftdef}
\end{equation}
which comes from the definition of vorticity. From all of the above one finds that the distorted velocity can finally be expressed as 
\begin{subequations}\label{eq:uwhom}
\begin{align}
&\hat{u}_1^{(H)}(t)=\left( 1 + \frac{\k_1 \k_3 \beta}{\k^2} \right) \hat{u}_{10}^{(H)}+ \frac{\k_1^2 \beta}{\k^2} \hat{u}_{30}^{(H)}, \label{uhom} \\
&\hat{u}_3^{(H)}(t) = \left( \frac{\k_0^2}{\k^2} - \frac{\k_1 \k_{30} \beta}{\k^2} \right) \hat{u}_{30}^{(H)} - \frac{\k_{12}^2 \beta}{\k^2} \hat{u}_{10}^{(H)}, \label{whom}
\end{align}
\end{subequations}
where $\k_{12}=(\k_1^2+\k_2^2)^{1/2}$, $\bk_0=(\k_{10},\k_{20},\k_{30})$ and $\k_0=|\bk_0|$, and we only focus on $\hat{u}_1$ and $\hat{u}_3$ because these are the velocity components necessary to calculate the shear stress.

The previous calculations only apply to the turbulence far away from the air-water interface, at least a distance $\sim L$ where $L$ is a characteristic integral length scale (but affected by the Stokes drift, because of the scale separation $L \ll \lambda$). In order to take into account the blocking effect of the air-water interface, where we assume that the interface affects the turbulence essentially as a frictionless wall for depths of $O(L)$ (because the air-water interface has a large density contrast), this effect can be accounted for by adding an irrotational correction to the homogeneous turbulent velocity field, as done before by \citet{Hunt_Graham_1978} and \citet{Teixeira2002}. Note that if the waves that generate the Stokes drift are irrotational (which is true to a good degree of approximation), then this irrotational correction remains irrotational. Hence, the turbulent velocity components affected both by the Stokes drift and by blocking can be expressed as two-dimensional Fourier integrals along the horizontal directions,
\begin{equation}
    u'_i(\boldsymbol{x},t) = \iint \hat{u}_i(\k_1,\k_2,z) e^{i(\k_1 x+ \k_2 y)} \rmd\k_1 \, \rmd\k_2,
    \label{ft2d}
\end{equation}
because the inhomogeneity imposed by blocking does not allow Fourier transformation in the vertical direction. 

Based on \citet{Hunt_Graham_1978} and \cite{Teixeira2002}, the Fourier transforms of $u'_1$ and $u'_3$ can be written
\begin{subequations}\label{uwboth}
\begin{align}
    &\hat{u}_1= \int \left( \hat{u}_1^{(H)} e^{i \k_3 z} - i \frac{\k_1}{\k_{12}} e^{\k_{12} z} \hat{u}_3^{(H)} \right) \rmd\k_3, \label{uboth} \\
    &\hat{u}_3 = \int \hat{u}_3^{(H)} \left( e^{i \k_3 z} - e^{\k_{12} z} \right) \rmd\k_3. \label{wboth}
\end{align}
\end{subequations}
For a flow associated with a surface wave, the blocking condition actually applies perpendicularly to the wavy air-water interface, which when adopting a model accounting for the individual wave cycles requires the use of a curvilinear coordinate system, as in \cite{Teixeira2002}. However, since our present model only includes the Stokes drift effect, no explicit wavy deformations of the interface are accounted for, and the surface is assumed to be in its average state, i.e., flat and horizontal at $z=0$. The solutions \eqref{uwboth} take this into account.

We wish to evaluate $\ouw = \ouwnum$ after some time of deformation. This can be done by using  (\ref{sols2}), (\ref{eq:uwhom}), (\ref{ft2d}), and also noting that, by definition
\begin{equation}
    \overline{\hat{u}_{i0}^{(H)}(\bk_0)^* \hat{u}_{j0}^{(H)}(\bk_0')} = \Phi_{ij}^{(H)}(\bk_0) \delta(\bk_0 - \bk_0'),
    \label{spec3d}
\end{equation}
where an asterisk denotes the complex conjugate,
$\Phi_{ij}^{(H)} (\bk_0)$ is the three-dimensional spectrum of the initial homogeneous and isotropic turbulence and $\delta$ is the three-dimensional Dirac delta, to show that $\ouwnum$ is given by
\begin{equation}
\overline{u'_1 u'_3} = \iiint \left( M_{13} \Phi_{13}^{(H)}+M_{11} \Phi_{11}^{(H)} + M_{33} \Phi_{33}^{(H)} \right) \rmd\k_1 \, \rmd\k_2 \, \rmd\k_{30},
\label{shearst}
\end{equation}
where, 
\begin{subequations}
\begin{align}
M_{13}=&\left( 1+ \frac{\k_1 \k_3 \beta}{\k^2} \right) \left( \frac{\k_0^2}{\k^2} - \frac{\k_1 \k_{30} \beta}{\k^2} \right) - \frac{\k_1^2 \k_{12}^2 \beta^2}{\k^4} \notag \\ 
&-\left[ \left( 1 + \frac{\k_1 \k_3 \beta}{\k^2} \right) \left( \frac{\k_0^2}{\k^2} - \frac{\k_1 \k_{30} \beta}{\k^2} \right)  - \frac{\k_1^2 \k_{12}^2 \beta^2}{\k^4} \right] e^{\k_{12} z} \cos( \k_3 z)  \nonumber \\ 
&+ 2 \frac{\k_1 \k_{12} \beta}{\k^2} \left( \frac{\k_0^2}{\k^2} - \frac{\k_1 \k_{30} \beta}{\k^2} \right) e^{\k_{12} z} \sin(\k_3 z), \label{defm13} \\
M_{11}=&-\frac{\k_{12}^2 \beta}{\k^2} \left[ 1 + \frac{\k_1 \k_3 \beta}{\k^2} - \left( 1 + 
\frac{\k_1 \k_3 \beta}{\k^2} \right) e^{\k_{12} z} \cos(\k_3 z)  \right.\nonumber \\
&\left. +\frac{\k_1 \k_{12} \beta}
{\k^2} e^{\k_{12} z} \sin(\k_3 z) \right], \label{defm11} \\
M_{33}=&\left( \frac{\k_0^2}{\k^2} - \frac{\k_1 \k_{30} \beta}{\k^2} \right) \left[ \frac{\k_1^2 \beta}{\k^2} \left( 1 - e^{\k_{12} z} \cos(\k_3 z) \right) 
\right.\notag \\
&\left.- \frac{\k_1}{\k_{12}}
\left( \frac{\k_0^2}{\k^2} - \frac{\k_1 \k_{30} \beta}{\k^2} \right) e^{\k_{12} z} 
\sin(\k_3 z) \right]. \label{defm33}
\end{align}
\end{subequations}

To calculate the shear stress using (\ref{shearst}), an expression for $\Phi_{ij}^{(H)}$ must be assumed. We employ the model \citep{Batchelor1953},
\begin{equation}
 \Phi_{ij}^{(H)}(\bk_0) = \left( \delta_{ij} - \frac{\k_{i0} \k_{j0}}{\k_0^2} \right) \frac{E(\k_0)}{4 \pi \k_0^2},
 \label{isotropic}
 \end{equation}
where $E(\k_0)$ is the energy spectrum of the homogeneous and isotropic turbulence, and since the calculations are inviscid, a suitable assumption for the energy spectrum is that first introduced by \citet{vonKarman1948}, expressed as
\begin{equation}
    E(\k_0)= \frac{g_2 q^2 L (\k_0 L)^4}{\left[ g_1 + (\k_0 L)^2 \right]^{17/6}},
  \label{vonkarman} 
\end{equation}
where $q$ is a characteristic r.m.s.\ value of a turbulent velocity component and $L$ is the longitudinal integral length scale. The constants $g_1 \approx 0.558$ and $g_2 \approx 1.196$ ensure that the definitions of $q$ and $L$ are consistent. 

Note that while a number of simplifying assumptions have been made which mean that full quantitative agreement with observations is not to be expected, the above theory involves no fitting or particular tailoring to our specific system. The turbulence is described 
by just two scalar parameters pertaining to the homogeneous bulk, $q$ and $L$, a crude simplification which nevertheless is expected to capture the most essential features. Importantly, these are quantities obtainable from point measurements in experiments as well as in the field.


\subsection{Input parameters}

Three quantities must be evaluated in order to obtain numerical values for $\ouw$ from \eqref{shearst}: $L, q$ and $\beta$.

The simplest to determine is $q$, which according to RDT satisfies 
\begin{equation}
    q^2 = {\textstyle \frac23}e,
\end{equation}
assuming isotropic turbulence, where the TKE $e$ is found in Table \ref{tab:mwt}.

The exact choice of turbulent integral length scale $L$ involves some amount of judgement. We shall take the most immediate available option and use the streamwise integral scale $L_x^x$ for $L$, reported for cases \iA--\iD\ by \cite{Smeltzer2023}. One does well to note that estimating the integral scale from experimental data is not trivial and several methods exist, which tend to yield significantly different results. The difference between the experiments means that no single method of calculating $L_x^x$ can be used for all, so direct quantitative comparison is dubious; 
however, RDT is only expected to describe the `spin-up phase' cases \iA--\iD\ from Experiment 1, which are directly comparable. 

The displacement $\beta$ defined in \eqref{eq:stokesdisp} is determined by the duration and extent of interaction between waves and turbulence in the time between generation by the active grid, and measurement a distance $\LFOV$ downstream. 
For a Gaussian wave packet like those in cases \iA--\iD, \citet{vandenbremer19} showed that the final value of $\beta$ after the passage of the group, $\betaf$ (subscript `$\mathrm{f}$': final), takes the form
\begin{equation}\label{eq:betagroup}
    \beta_\mathrm{f}(z) = 4 \sqrt{\pi} \sigma_0 k_0 (\epzp)^2\rme^{2 k_0 z},
\end{equation}
where $\sigma_0$ is the spatial standard deviation of the Gaussian wave group and $\epzp$ the peak steepness. By definition of wave group, 
$\sigma_0= c_g(k_0) \tau_0$ (see \eqref{eq:disprel} and Table \ref{tab:mwt}). The expression presumes that the turbulence moving downstream from the active grid has interacted with the whole wave group before reaching the field of view; this is true in our Experiment 1.

It might be interesting to also calculate the values of $\betaf$ for the cases with regular waves, which gives an indication of the extent of interaction between waves and turbulence before the turbulent flow reaches the field of view. 
The cases \iE--\iH\ consider regular waves so that the interaction occurs at a near-constant rate during the travel time $t_\mathrm{travel}$, hence in this case we may approximate
\begin{equation}\label{eq:betaregular}
    \beta = t_\mathrm{travel}\frac{\rmd \us}{\rmd z} = \frac{2k_0\LFOV}{U_0}(\epz)^2c(k_0)\rme^{2 k_0 z}.
\end{equation}
Values of $\betafz$ for all cases are given in Table \ref{tab:RDTparams}, and vary by more than an order of magnitude from the longest waves of lowest steepness to the short and steep waves.

\subsection{The RDT results}

\begin{table}
  \begin{center}
\def~{\hphantom{0}}
   \begin{tabular}{cccc}
      Case  &$\Tint$ & $\Tint/\vpz$ &  $\betafz$\\[3pt]
      &$(\mathrm{s})$ & & \\
      \hline      \iA  & $4.3$ & $6.6$  &$3.3$\\
      \iB  & $4.6$ & $7.0$  &$3.5$\\
      \iC  & $5.1$ & $7.6$  &$2.2, 4.7$\\
      \iD  & $5.0$ & $7.6$  &$4.6$\\
      \iE.1 & $29.2$ & $35.7$            & $3.6, 8.6, 14.2$\\
      \iE.2 & $29.2$ & $50.6$            &$13.9, 22.3, 27.2$\\
      \iF.1 &$29.2$  & $35.9$            & $2.1, 6.3, 12.7$\\
      \iF.2 & $29.2$ & $50.4$            & $12.1, 17.8, 27.1$\\
      \iG & $46.1$ & $82.5$ &$12.2,32.6$\\
      \iH & $46.1$ & $82.8$ &$12.2,32.7$\\
  \end{tabular}
  \caption{
      Derived wave quantities for use in RDT analysis.   For cases \iA--\iD, $\Tint=\sqrt{\pi}\tau_0$ is used with $\tau_0$ from Eq.~\eqref{eq:intrinsicwidth}, while for cases \iE--\iH, $\Tint=\LFOV/U_0$; $\betafz$ is found from Eqs.\ \eqref{eq:betagroup} and \eqref{eq:betaregular} for groups and regular waves, respectively. Note that it is related to $\Tint$ via Eq.~\eqref{eq:betaTint}.
  }
  \label{tab:RDTparams}
  \end{center}
\end{table}

\begin{figure}
  \centering
    \includegraphics[width=\textwidth]{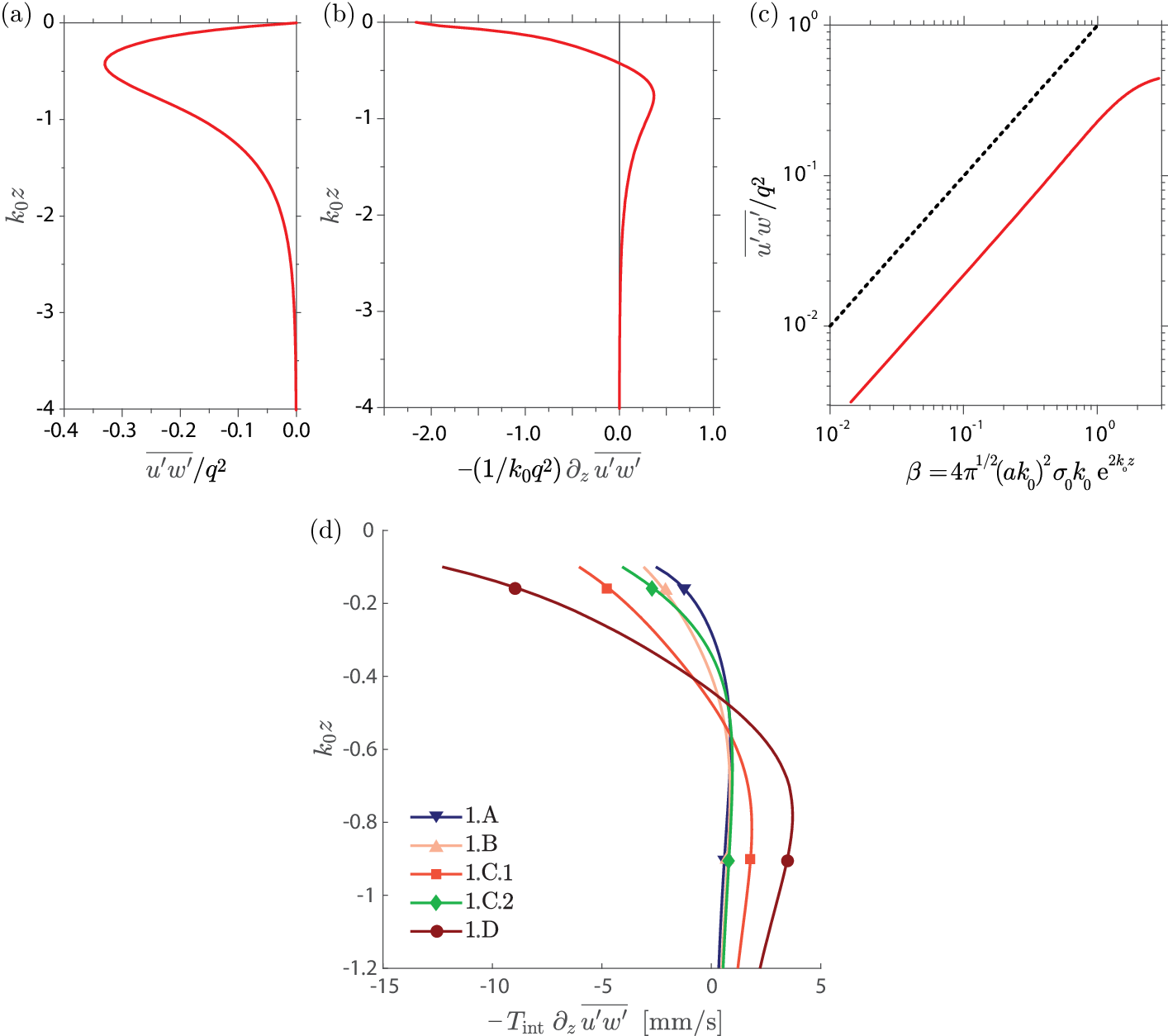}
  \caption{
    Results from RDT: (a) profile of the normalised shear stress $\ouw/q^2$   as a function of depth $k_0 z$; (b) profile of the normalized vertical derivative of the shear stress; (c) the normalised shear stress as a function of $\beta$ for $k_0 z=-0.4335$, the depth where $\ouw/q^2$ attains its maximum magnitude. The dashed line is the 1:1 line, illustrating a linear dependence. (d) The RDT estimates of the anti-Stokes velocity profile after the passage of the wave groups, corresponding to figure \ref{fig:wavegroups}(b).
  }
  \label{fig:RDT}
\end{figure}

Figure \ref{fig:RDT}(a) shows the profile of the Reynolds shear stress provided by RDT after passage of the Gaussian wave group, using the input parameters specified above. Figure \ref{fig:RDT}(b) shows the vertical derivative of this shear stress, which is proportional to the forcing of the mean current, according to (\ref{eq:RANS2}). The results are expected to represent a similar situation to our experiment 1 where wave groups were employed and the flow is still in a state of transition when the wave group has passed. One may also note the similarity in shape with the measured currents also for regular waves presented in figure \ref{fig:delU_lin}, indicative that the same process has been at play.

Next, figure \ref{fig:RDT}(c) shows the dependence of the shear stress at its maximum on $\beta$ (which is proportional to the square of the wave slope $\epz$ as Eqs.\ \eqref{eq:betagroup} and \eqref{eq:betaregular} show). The vertical derivative of the shear stress at the surface, to which the current at the surface must be proportional, is clearly proportional to this maximum. This implies that the current at the surface (departing from a quiescent flow) must be proportional to $(\epz)^2$, which is consistent with the results of figure \ref{fig:delUmax} for cases \iE\ and \iF. 

A shear stress profile such as depicted in figure~\ref{fig:RDT}(a) would, according to (\ref{eq:RANS2}), lead to a current intensifying continuously and indefinitely in time. However, the current will generate a shear stress of opposite sign to that associated with the Stokes drift, and when these shear stresses have evolved so that they cancel each other, the current will reach a steady state (see \cite{Pearson_2018}). Moreover, dissipation, which has been neglected in the present RDT treatment, is always present, and would act in a similar way to limit the current growth.

\subsubsection{Effective interaction time}

While the primary goal of our RDT model is to elucidate the physical process behind our observations, we have also used \eqref{shearst} to obtain a rough estimate of the current resulting from the wave groups in Experiment 1 (Cases \iA--\iD) encountering the incoming turbulence. 
In order to obtain quantitative values, we use a simplified procedure: noting that $\p_z \ouw \propto (\epz)^2$, we estimate the current after the passage of the group as
\begin{align}
    U_\text{RDT}(z) &\approx -\int \partial_z \ouw(t) \rmd t 
    \approx -(\p_z \ouw)_\text{RDT}\int_{-\infty}^\infty 
    \frac{\tilde{a}(t)^2}{a^2_\mathrm{p}}
    \rmd t\notag \\
    &\approx -(\p_z \ouw)_\text{RDT}\Tintgrp \label{eq:URDT}
\end{align}
where we have introduced an ``effective interaction time" for a wave group,
\begin{equation}\label{eq:Tint}
  \Tintgrp = \frac1{\ap^{2}}\int_{-\infty}^\infty \tilde{a}(t)^2\rmd t,
\end{equation}
an estimate for the duration for which the turbulent current has been influenced by the waves before it reaches the point where it is measured, taking into account that the wave amplitude varies throughout the group. As seen by the turbulence, $\tilde{a}(t)$ is a Gaussian with $\tau_0$ from Eq.~\eqref{eq:intrinsicwidth} replacing $\tau$ in \eqref{eq:wamp}, inserting which into \eqref{eq:Tint} gives
\begin{equation}
    \Tintgrp = \sqrt{\pi}\tau_0.
\end{equation}
We next insert into \eqref{eq:URDT}
the calculated Reynolds stress vertical gradient shown in figure \ref{fig:RDT}(b) (multiplied by $k_0q^2$ to obtain a dimensional value), using values for $k_0, q=2e/3,L=L_x^x$ and $\Tintgrp$ from table \ref{tab:RDTparams}. The resulting estimated velocity profiles are shown in figure \ref{fig:RDT}(\id). 

It is worth noting the relation between the effective interaction time and the fluid displacement $\beta$. For regular waves, the turbulence interacts constantly with waves throughout its passage from grid to field of view, so the interaction time in the presence of regular waves is $\Tintreg = \LFOV/U_0$. Using Eq.~\eqref{eq:betagroup} and noting that for deep-water waves, $\sigma_0=c_g\tau_0=\half c\tau_0$, we find that for regular waves as well as wave groups it holds that
\begin{equation}\label{eq:betaTint}
    \beta(z) = 2k_0 \Tint u_s(z).
\end{equation}

\subsubsection{Comparison with experiment}

There is little reason to expect close quantitative agreement between our measured $\Delta U(z)$ in Experiment 1 and $U_\text{RDT}(z)$ because the assumptions behind \eqref{eq:URDT} can only hold very early in the `spin-up' period; the relation assumes that $\ouw$ retains the vertical shape shown in figure \ref{fig:RDT}(a) throughout the duration of the interaction between wave group and turbulence. Gradually, however, the turbulent flow field begins to `push back' via the growth of other terms in \eqref{eq:uwtransp} until $\partial_z\ouw$ eventually becomes very small. Investigating this transient process in detail is a highly relevant question for future study, experimentally and numerically \cite[e.g.\ in the vein of][]{Guo2013}.

Despite the na\"{i}vet\'{e} of \eqref{eq:URDT} and the numerous assumptions made when applying RDT, the agreement with the measurements of $\Delta U$ in figure \ref{fig:wavegroups}(b) is not only qualitatively but also quantitatively reasonable. The predictions vary with depth notably faster than the measured currents, which we conjecture is a consequence of the smoothing effect of turbulence as it develops beyond the linear-growth regime described by \eqref{eq:URDT}, a
turbulent mixing which RDT by construction cannot account for.
Agreement is comparatively poor in Case \iB\ where $U_\text{RDT}$ is far smaller than the corresponding $\Delta U$, which we cannot explain at present (note that this same case displays a surprisingly strong modification of the underlying turbulence considering that the TKE of Case \iB\ is intermediate, as seen in figure 4 of \cite{Smeltzer2023}). 

The key conclusion to be drawn is perhaps that the change in Eulerian current can indeed be ascribed to the interaction between waves and turbulence, to wit, the gradual tilting and stretching of vortices by the Stokes drift, as previously studied by \cite{Teixeira2002}. 

\section{Conclusions}

We have presented experimental evidence that an Eulerian-mean flow directed opposite to the waves' propagation direction is created near the water surface when waves propagate atop a turbulent flow, and argued via two different theoretical approaches how the current is the result of waves and turbulence interacting. 

Three experiments were conducted, all including waves propagating upstream on initially depth-uniform flows with different turbulent properties. Bespoke turbulence was created with an active grid and velocity fields were measured with PIV. One experiment compared flow conditions before and after the passage of groups of waves, while two studied the mean flow under regular waves, one with low acquisition frequency over a long time, the other at higher frequency in repeated intervals.  

Experiments as well as theory show how, when irrotational waves and a turbulent current encounter each other, the combined flow goes through a period of transition until a new quasi-steady state is reached. The fundamental mechanism involved is the rearrangement of horizontal momentum driven by the Reynolds stress which arises when the Stokes drift acts on turbulent eddies, as studied, e.g., by \citet{Teixeira2002}.

Our experiment with groups of waves studies both the transition period and the final equilibrium state. The former is investigated by allowing the turbulence a limited time to interact with passing wave group. In the two experiments involving regular waves, on the other hand, the final quasi-steady state appears to have been reached by the time the flow is measured.

We present two separate theoretical models which can describe the quasi-steady state and the transition period, respectively.
For the latter situation an approximate relation between the mean current shear $\rmd \bu/\rmd z$ and the Stokes drift gradient $\rmd \us/\rmd z$ is derived following \citet{Pearson_2018}, valid nearest the surface where $\us$ is significant. The relation involves the turbulent variances $\ouu$ and $\oww$ and shows good enough agreement with experiments to inspire confidence in its use, in the cases where underlying assumptions are satisfied.

A further approximate relation is adapted from \citet{Pearson_2018} and relates the rate of change of the Eulerian-mean current to the Reynolds stress $\partial_z\ouw$. Hypothesising that the underlying mechanism is the action of Stokes drift on turbulent eddies, Rapid Distortion Theory (RDT) is used to estimate $\ouw(z)$ and hence the time-varying $\bu(z,t)$ in the early phase of interaction. The predictions of RDT are compared with the measurements of $\bu(z)$ due to passing wave groups, by using measured values of TKE, integral length scale, and group envelope width. Qualitative and quantitative agreement with measurements is found, despite a series of simplifying assumptions.

\section*{Acknowledgements}

We have benefited from discussions with a number of colleagues, particularly Victor Shrira and Nick Pizzo. We thank Marc Buckley for invaluable input and for sharing data-analysis code, and Masoud Asadi, Leon Li and Pim Bullee for assistance with the experiments. 
We are grateful to Rafael B\"{o}lsterli for feedback on the manuscript.

\section*{Funding}
The work was co-funded by the Research Council of Norway (\emph{iMOD}, 325114) and the European Union (ERC StG, \emph{GLITR}, 101041000 and ERC CoG, \emph{WaTurSheD}, 101045299). 
Views and opinions expressed are, however, those of the authors only and do not necessarily reflect those of the European Union or the European Research Council. Neither the European Union nor the granting authority can be held responsible for them. 
M.A.C.T.\ was supported by Portuguese national funds through FCT/MCTES (PIDDAC), for the ResearchUnits CEFT:UID/00532, and ALiCE: LA/P/0045/2020 (doi: 10.54499/LA/P/0045/2020).

\section*{Declaration of interest}
The authors report no conflict of interest.

\section*{Author contributions}
Primary contributions are as follows. 
S.{\AA}.E.: Principal writer, theory, project co-ordination, supervision.
O.R.: Planned and performed experiment 3, conducted the majority of the data analysis. 
B.K.S.: Planned and performed experiments 1 and 2, first discovered the change in current in the experimental data. 
M.A.C.T.: Developed the Rapid Distortion Theory. 
T.S.vdB.: Theory development, discussions. 
K.S.M.: Error and convergence analysis. 
R.J.H.: Supervision of experiments, turbulence-theoretic contributions and discussion. 
S.{\AA}.E. and O.R.\ should be considered to have made equal contributions.
All authors contributed significantly to the writing of the manuscript. 

\section*{Data availability statement}
The experimental data from experiment 1 are available at \href{https://doi.org/10.18710/R0I0RW}{doi:10.18710/R0I0RW}, and for experiments 2 and 3 from \href{https://doi.org/10.18710/ZZPUNJ}{doi:10.18710/ZZPUNJ}.

\appendix


\section{Active grid settings} \label{app:grid}

The settings used for the active grid in the different cases $\iA$ -- $\iH$ are listed in table \ref{tab:AGprotocols}. The active grid contains two sets of bars; vertical and horizontal. The phrase \textit{static, open} indicates that the grid bars are not moving but in a fully open position, causing as little blockage of the inflow as possible. Cases labelled \textit{random rotation} means that grid bars are actuated randomly with a top-hat distribution centred at $\overline{f_G}$ spread over $\pm \delta f_G$. The rotation direction is also random with an equal likelihood of clockwise and counter-clockwise rotation. For case $\iH$ the grid bars are flapped $\pm 60^\circ$ about the fully open position. Here, a flap motion occurs at random intervals uniformly distributed between 
$0.5$ and $1.0$ seconds. Each bar is flapped independently.

\begin{table}
    \centering
    \begin{tabular}{ccll}
         Case & $\overline{f_G} \pm \delta f_G$ [Hz] & Horizontal bar & Vertical bars  \\
         \hline
         \iA & $0$ & static, open & static, open \\
         \iB & $1.5 \pm 0.75$ & static, open & random rotation \\         
         \iC & $1.5 \pm 0.75$ & random rotation & random rotation \\         
         \iD & $0.2 \pm 0.1$ & random rotation & random rotation \\         
         \iE & $0.05 \pm 0.025$ & random rotation & random rotation \\         
         \iF & $1.0 \pm 0.5$ & random rotation & random rotation \\         
         \iG & $1.5 \pm 0.75$ & static, open & random rotation \\         
         \iH & - & static, open & random flapping $\pm 60^\circ$          
    \end{tabular}
    \caption{Active grid protocols used. }
    \label{tab:AGprotocols}
\end{table}


\section{Errors and convergence in experimental current measurements}\label{app:errors}

The wave-induced current profiles, $\Delta U$, shown in figures \ref{fig:wavegroups} and \ref{fig:delU_lin} are one to two orders of magnitude smaller than the mean currents themselves, and a careful analysis of errors and convergence must be conducted to evaluate significance and accuracy. An error in the measured average velocity profiles must be well below $1$\,mm/s. While this is less than the uncertainty of a single measurement, errors in average values can be far smaller. 

\begin{figure}
    \centering
    \includegraphics[width=0.98\linewidth]{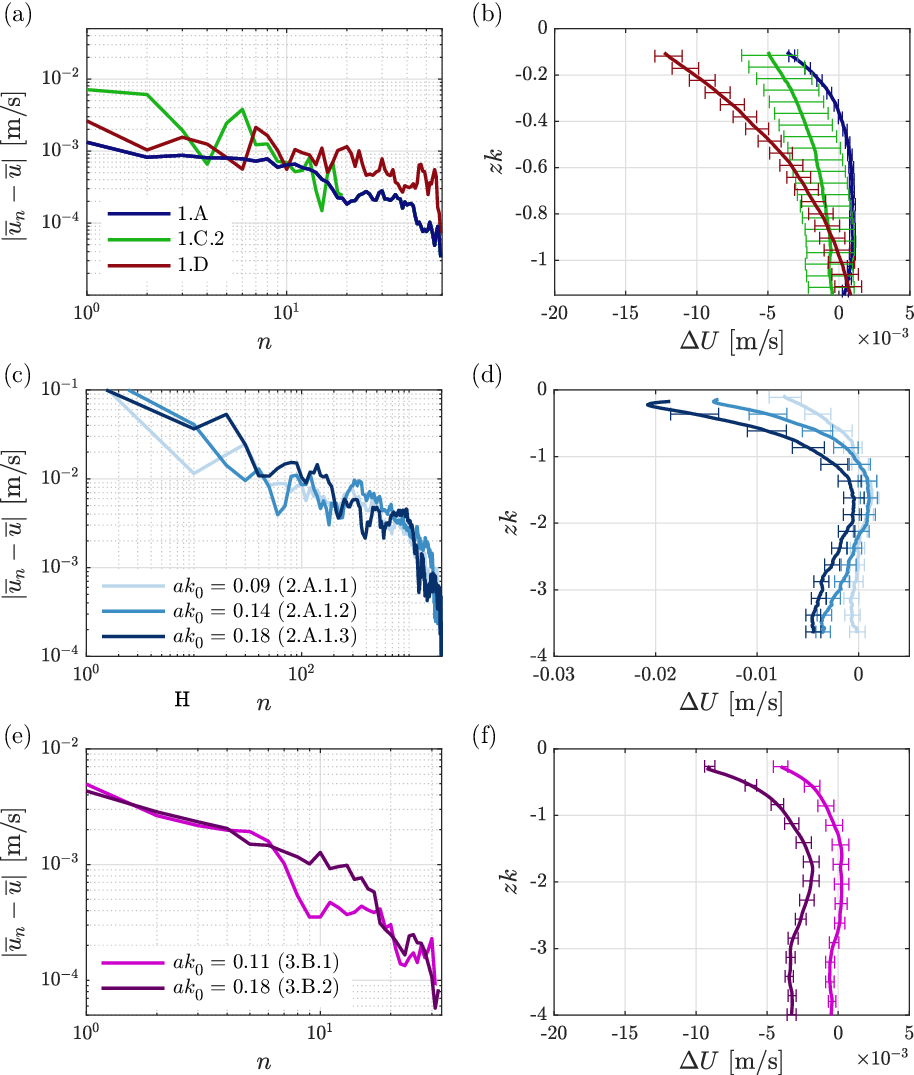} 
    \caption{(a,c,e) Convergence of mean velocities for the representative cases indicated in the legends, for increasing number of ensembles (a,e) or snapshots (c). (b,d,f) Velocity profiles for the same cases with error bars indicating the standard deviation from $2000$ bootstrapped profiles.
    }
    \label{fig:Conv_and_uncertainty_AH}
\end{figure}

Convergence of the velocity measurements is presented in the left-hand panels of figure \ref{fig:Conv_and_uncertainty_AH}. Because the results are highly similar for all cases within each experiment, we only show three cases from Experiment 1 (\iA, \iC.2 and \iD), and one each from Experiments 2 and 3 (\iE.1 and \iH, respectively). The absolute difference in velocity between the average of $n$ measurements, $\bau_n$, and that of all measurements, $\bau$, is shown for the depth-averaged mean velocity.
In Experiment 1 very little wave motion was present during the measurement time intervals, hence the fluctuations are mainly due to turbulent motion. Case \iA\ has the lowest turbulence level and the average varies very little even for small $n$, while also for the most turbulent of our cases, \iD, the average shows variations well below $1$\,mm/s for $n\gtrsim 30$ (out of the total of $60$). For the cases where $60$ repetitions were made (i.e., all except \iC.2) the result is well converged to better than $\sim 10^{-4}$\,m. Fewer repetitions were performed for case \iC.2, but convergence is also sufficiently good so that further repetitions would not significantly change the curve plotted in figure \ref{fig:wavegroups}. 

It is particularly pertinent to check for convergence in Experiments 2 and 3, where waves are present during the measurements, with instantaneous orbital velocities which cause differences from frame to frame which far exceed $\Delta U$. In Experiment 2 individual snapshots of the velocity field were taken at low frequency. In figure \ref{fig:Conv_and_uncertainty_AH}(c) we show the absolute difference (L$_1$ norm) between the average over the first $n$ snapshots, and the full set of $2000$ snapshots. Although convergence is slow as can be expected, averages vary by less than $1$\,mm/s after approximately $n=1200$ frames, and it is clear that a longer time series would not significantly alter the results. Likewise, convergence with increasing number of repeated measurements in Experiment 3 is shown for case \iH, representative also of case \iG. The average no longer fluctuates significantly after approximately $20$ out of the $32$ 
$25$-second intervals.

The right-hand panels of figure \ref{fig:Conv_and_uncertainty_AH} show the velocity profiles for the same cases, indicating the standard deviations calculated using the bootstrapping method.
This method was introduced by \cite{Efron1979} and applied to turbulence research by \cite{Benedict1996}. 
Velocity profiles were calculated by averaging over $n_\text{tot}$ randomly selected individual measurements (including repetitions) where $n_\text{tot}$ is the total number of measurements; this was done $2000$ times and the standard deviation was found and shown as error bars. As might be expected, the greatest uncertainty is for case \iC.2 where fewer repetitions were made. It is the least certain of our measurements, though it is noteworthy that it follows the same trend as the other cases, and the values of $\Delta U$ are clearly significant. We note, as expected, that for Experiment 1 the standard deviation is nearly constant with depth, while for the cases where waves were present, uncertainties are higher near the surface. Particularly in Experiment 2, the standard deviation of measurements near the surface are several mm/s, but far smaller than the measured value of $\Delta U$.


\section{Separating waves from turbulence}\label{app:POD}

\begin{figure}
\centering

    \begin{subfigure}[t]{0.3\textwidth}
        \centering
        \includegraphics[width=\textwidth]{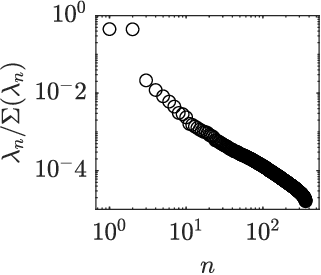}
        \subcaption{}
    \end{subfigure}
    \begin{subfigure}[t]{0.3\textwidth}
        \centering
        \includegraphics[width=\textwidth]{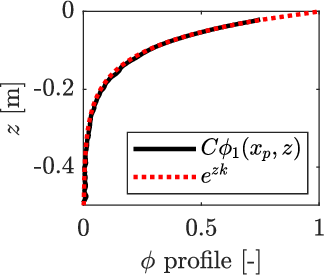}
        \subcaption{}
    \end{subfigure}
    \begin{subfigure}[t]{0.3\textwidth}
        \centering
        \includegraphics[width=\textwidth]{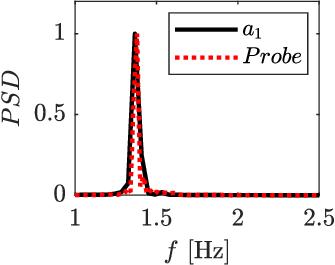}
        \subcaption{}
    \end{subfigure}

    \vspace{0.5em}
    
    \begin{subfigure}[t]{0.9\textwidth}
        \centering
        \includegraphics[width=\textwidth]{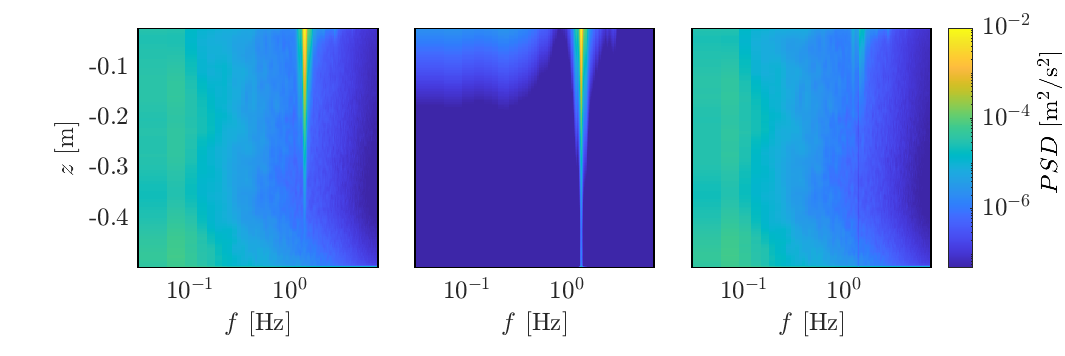}
        \subcaption{}
        
    \end{subfigure}

    \vspace{0.5em}
    
    \hspace{1.5em}\begin{subfigure}[t]{0.86\textwidth}
        \centering
        \includegraphics[width=\textwidth]{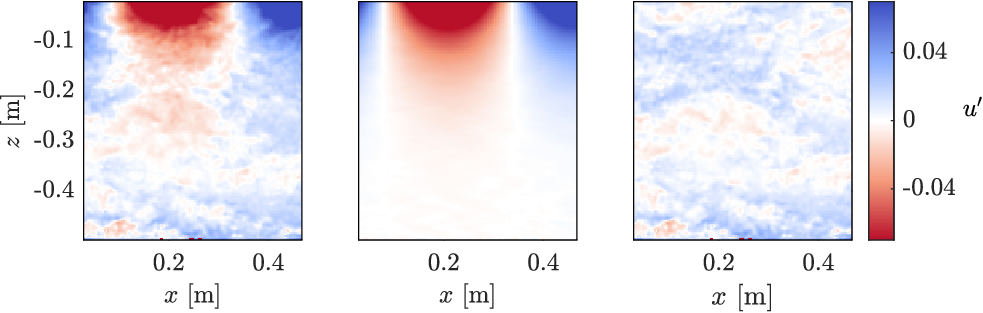}
        \subcaption{}
    \end{subfigure}
    
    \caption{
    wave--turbulence decomposition of the streamwise mean-subtracted velocity field $\uwt$ using POD.
    (a) Normalised mode energy $\lambda_n$ for a single ensemble of case \iG.2. (b) Vertical slice of mode 1 beneath a wave peak ($x=x_\mathrm{p}$). A factor $C$ is used for normalisation. (c) Power spectral density (PSD) of the temporal coefficients of mode 1 and the wave probe signal. (d) Spectrogram of the streamwise velocity at each depth coordinate; see equation  \eqref{eq:triple}. (e) Snapshots of the decomposed signal. From left to right in panels (d) and (e):     $\uwt$, $\tu$, and $u'$. A supplementary movie illustrating wave--current decomposition of our data is available at \url{https://doi.org/10.1017/jfm.2026.11163}.
    } 
    \label{fig:Appendix_POD}
\end{figure}

For the analysis in figure \ref{fig:StokesDriftVsEuler_EH} a triple decomposition according to \eqref{eq:triple} is needed. 
The task is nontrivial as demonstrated by many authors in the past \citep[see the thorough review by][]{Chavez-Dorado2025}, and particularly challenging for our cases \iG\ and \iH\ where phase information is not available. Standard methods such as phase-conditioned averaging (PhCA), and other methods such as Dynamic Mode Decomposition \citep{Chavez-Dorado2025}, Synchrosqueeze Wavelet Transform \citep{Perez20} and Empirical Mode Decomposition \citep{Peruzzi2021} cannot be employed
for data from Experiment 2. In contrast, measurements from Experiment 3 have a spatially similar plane of measurement and field of view, and are resolved in time so that both PhCA and POD can be used, allowing us to validate POD for cases \iG\ and \iH. 

We find that the method of Proper Orthogonal Decomposition (POD) \citep{Berkooz1993, Taira2017} is highly effective for this purpose even when only individual images of the turbulent field are available. We quantify this in the following and compare with the more standard PhCA for cases \iE\ and \iF.

\subsection{Proper Orthogonal Decomposition}

Here we briefly present the principle of POD while referring to specialised literature for details \citep[e.g.,][]{Berkooz1993, Taira2017}. In short, after subtracting the mean velocity $\bau(z)$, each component of the remaining field which we denote $\buwt=\bu-\babu=\tbu + \bu'$ is decomposed into $N$ orthogonal spatial modes, $\phi_n(\mathbf{x})$, and their respective temporal coefficients, $a_n(t)$, as
\begin{equation}\label{eq:POD}
    \uwt(\mathbf{x}, t) = \sum_{n=1}^Na_n(t)\phi_n(\mathbf{x})
\end{equation} 
($u$ is understood to be either velocity component) where $N$ is the number of measured PIV fields, and each mode $n$ has an associated `energy' $\lambda_n$. We calculate the POD using the method of snapshots. 

Figure \ref{fig:Appendix_POD} illustrates POD performed on case \iG. Panel (a) shows the POD mode energy distribution for a single ensemble. Note how the two highest-energy modes, $\lambda_1$ and $\lambda_2$ contribute approximately $90\%$ of the energy in the flow indicating that POD identifies a low-rank structure \citep[see, e.g.,][]{Taira2017}. It transpires that according to criteria we will return to, these two modes can be identified as containing the wave motion $\tbu$. 

Qualitative checks of the correspondence between the two first POD modes and wave motion are shown in figure \ref{fig:Appendix_POD}(b) and (c), the former demonstrates adherence of $\phi_1(z)$ at the peak position $x=x_\mathrm{p}$, to the exponential depth dependence of potential waves at the carrier frequency; the latter shows how the temporal coefficient $a_1(t)$ follows the same Gaussian spectrum as that measured by a wave probe at the free surface. Mode $\lambda_2$ is identical to $\lambda_1$ except for a phase shift by an angle $\pi/2$, thus for panel (c) we only show the first mode. 
The wave-only velocity field is now taken to be $\tu(\mathbf{x}, t)=\sum_{n=1}^2a_n(t)\phi_n(\mathbf{x})$, and the turbulent field is $u'(\mathbf{x}, t)=\sum_{n=3}^Na_n(t)\phi_n(\mathbf{x})$. 

A more quantitative test of the method's performance is Figure \ref{fig:Appendix_POD} (d) which shows a depth-wise power spectral density. By calculating the Fourier transform in time for points at the same $z$, the plot visualises where in the water column, and for which frequencies, kinetic energy has been extracted. The peak in energy around the wave frequency is nearly exclusively present in the spectrum of $\tu(z)$ (central panel). A small amount of energy is present also at frequencies far below $\omega_0$, which is expected due to minor fluctuations in the mean velocity, subharmonic waves, and low-frequency sloshing modes in the tank. 
Finally, for illustrative purposes, a decomposed signal from a single PIV frame is displayed in Figure \ref{fig:Appendix_POD}(e).

\subsection{Phase-conditioned average}

\begin{figure}
\centering

    \begin{subfigure}[t]{0.6\textwidth}
        \centering
        \includegraphics[width=\textwidth]{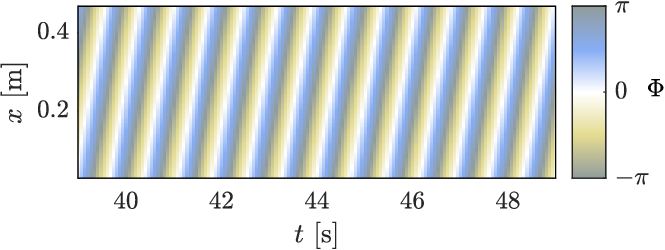}
        \subcaption{}
    \end{subfigure}
    \begin{subfigure}[t]{0.3\textwidth}
        \centering
        \includegraphics[width=\textwidth]{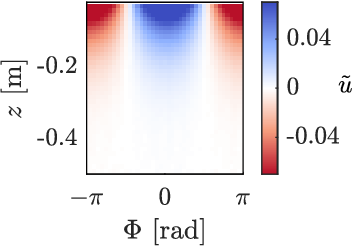}
        \subcaption{}
    \end{subfigure}
    
    \vspace{0.5em}
    
    \begin{subfigure}[t]{0.92\textwidth}
        \centering
        \includegraphics[width=\textwidth]{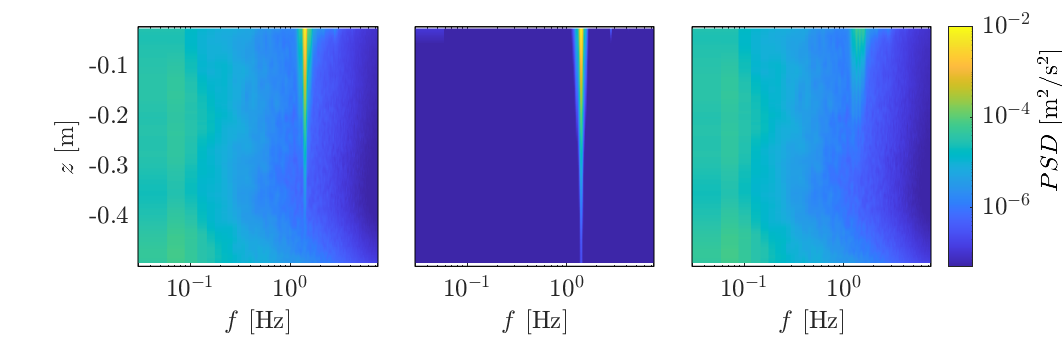}
        \subcaption{}
    \end{subfigure}
    
    \vspace{0.5em}
    
    \hspace{1.5em}\begin{subfigure}[t]{.95\textwidth}
        \centering
        \includegraphics[width=\textwidth]{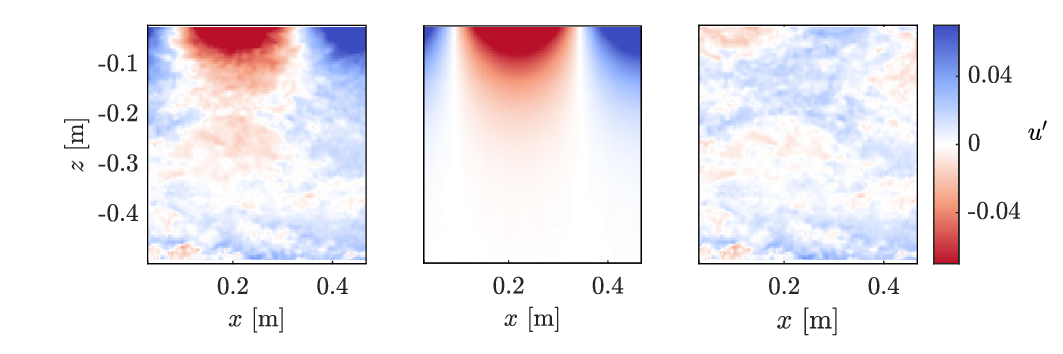}
        \subcaption{}
    \end{subfigure}
    \caption{
    wave--turbulence decomposition  of the streamwise velocity component  using PhCA from measurements of case \iG.2.  (a) Sample of the wave phase field, $\Phi(x, t)$, showing the phase variation across the spatial domain just beneath the wave trough. 
    (\ib) Phase-resolved average velocity, $\uwph(\Phi, z)$. Panels (c) and (d) show the same as figures \ref{fig:Appendix_POD}(d) and (e), respectively, for PhCA instead of POD.}
    \label{fig:Appendix_PhCA}
\end{figure}

Variations of the PhCA method have been employed in wave/turbulence research for a long time. 
Following the procedure of \citep{Buckley2017}, the wave motion is obtained from $\uwt$ by extracting the mean velocity field 
for each value of the depth and the phase of the wave, denoted $\uwph(\Phi, z)$. 

The wave phase, $\Phi(x, t)$, is extracted from the analytic signal of the velocity at a depth layer just beneath the wave trough. At each position, the analytic signal is acquired from the Hilbert transform in the temporal direction \citep{Melville1983}. A sample wave phase field is depicted in \ref{fig:Appendix_PhCA}(a). The phase is divided into bins, and the phase-conditioned average, $\uwph(\Phi, z)$ is calculated by taking the average of the mean-subtracted velocity $\uwt(\bx, t)$ in each phase bin. A sample of the resulting phase-resolved average, $\uwph(\Phi, z)$, is shown in Figure \ref{fig:Appendix_PhCA}(b).
The wave component of the velocity field is taken to be $\tu (\mathbf{x}, t) =\uwph(\Phi(x,t),z)$. Figure \ref{fig:Appendix_PhCA}(c) shows the power spectral density as a function of depth for comparison with figure \ref{fig:Appendix_POD}(d), and figure \ref{fig:Appendix_PhCA}(d) shows the same snapshot of the velocity field as figure \ref{fig:Appendix_PhCA}(e).

\begin{figure}
    \centering
    \includegraphics[width=\linewidth]{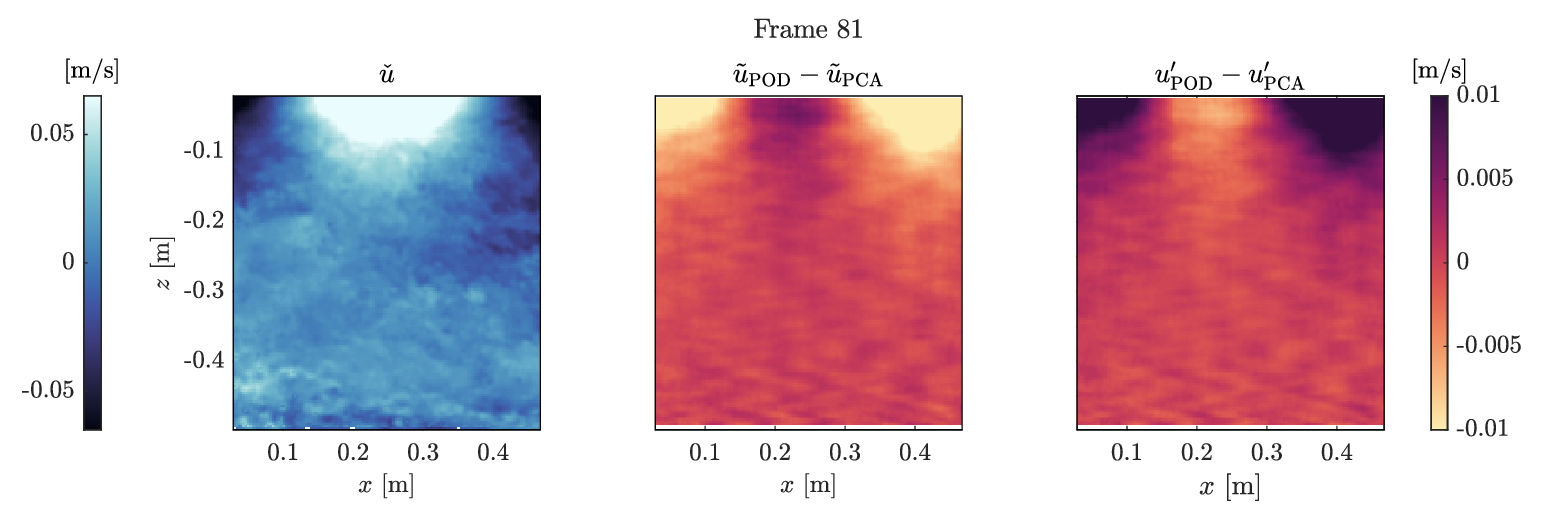} \\
    \includegraphics[width=\linewidth]{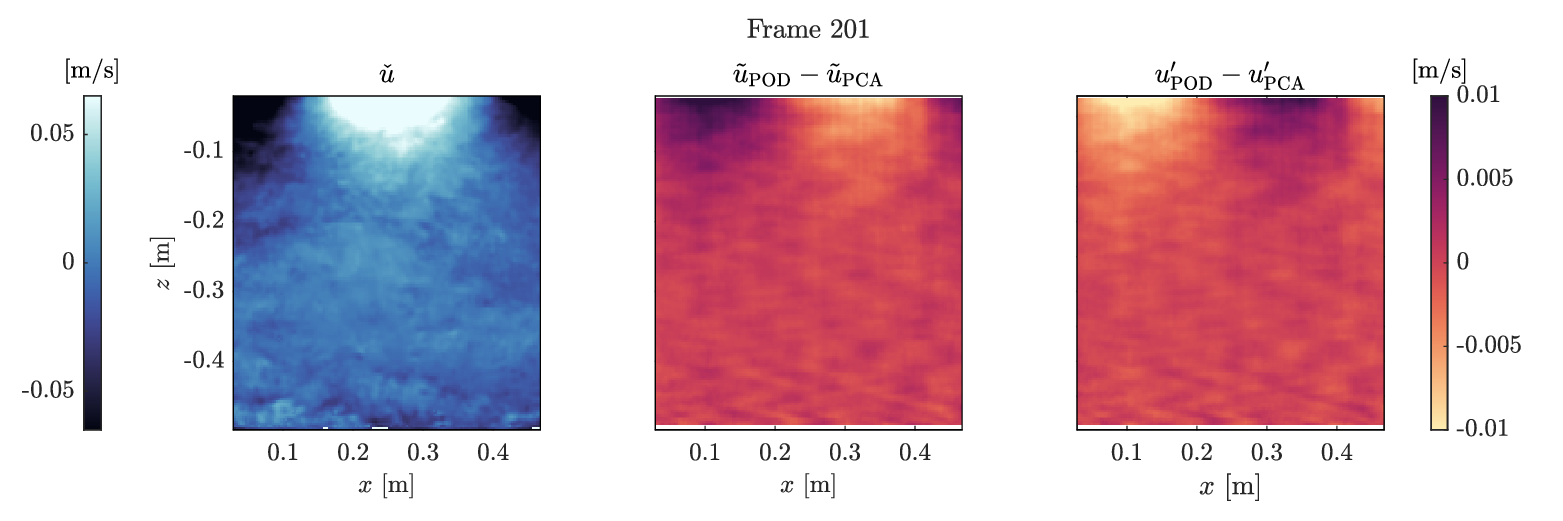} 
    \caption{Difference between triple-decomposed streamwise velocity fields for two different snapshots with approximately the same phase from case \iG.2, streamwise velocity component (waves moving left to right, current moving right to left). Left panels: full mean-subtracted velocity field $\uwt$; middle panels: difference in wave velocities; right panels: difference in turbulent velocities.
    Top row: error due to PhCA amplitude error; bottom row: error due to PhCA phase error.
    }
    \label{fig:decompDiff}
\end{figure}

At a qualitative level, both methods do well in separating waves from turbulence in our data. Because wave velocities are far higher than turbulent velocities near the surface, however, ascribing even a small percentage of the wave motion to $u'$ and $w'$ could have a significant effect on the variances $\ouu$ and $\oww$. 

We compare first the qualitative information in the snapshots of the decomposed velocity fields. In the case of POD in figure \ref{fig:Appendix_POD}(e), some tiny velocity fluctuations remain in the supposed wave component, faintly visible in the lower half of the central panel. These fluctuations are not correlated to the turbulent flow field and are an inescapable artefact of the mode decomposition procedure. PhCA by construction produces a field $\tbu$ which is smooth, as seen in figure \ref{fig:Appendix_PhCA}(d). The snapshots of the extracted $u'$ in the rightmost panels  of figures \ref{fig:Appendix_POD}(e) and \ref{fig:Appendix_PhCA}(d) are hardly possible to distinguish by eye, but the quantitative comparison shows that more energy is ascribed to turbulence (and less to waves) for PhCA than POD. 

An even closer comparison reveals that there are two main types of discrepancies between the two methods. First, the wave velocity $\tu$ also contains considerable signal at frequencies well below $\omega_0$; surface waves of such low frequencies are considerably longer than the field of view and PhCA cannot detect them. 
The second difference is clearly seen when regarding in the depthwise spectra of $u'$
in figures \ref{fig:Appendix_POD}(d) and \ref{fig:Appendix_PhCA}(c), when we consider the frequencies near the carrier-wave frequency $f_0$ (equal to $1.78$~Hz in this example) where the spectrum of the whole field $\uwt$ has a very pronounced peak.  
Both methods leave some residue of the wave in the turbulent signal after waves have been removed, but it is significantly smaller for POD than for PhCA, indicating that the latter leaves a larger `wave residue' in the identified turbulence field $\bu'$. PhCA produces a more narrowband $\tbu$ with hardly any signal outside of this peak, whereas the POD wave velocities produce a broader wave spectrum. 

A look at the difference between the POD and PhCA wave fields, as in the middle panels in figure \ref{fig:decompDiff}, reveals that the wave-signal assigned to turbulence is due to slight inaccuracies of PhCA to determine phase and amplitude. The top row (Frame 81) shows an amplitude mismatch (the difference field is in phase with the wave), and the bottom row (Frame 201) shows a phase mismatch (difference field approximately $\pi/2$ out of phase with the wave).


\section{Reynolds stress measurements}\label{app:Reynolds}

\begin{figure}
    \centering
    \includegraphics[width=\linewidth]{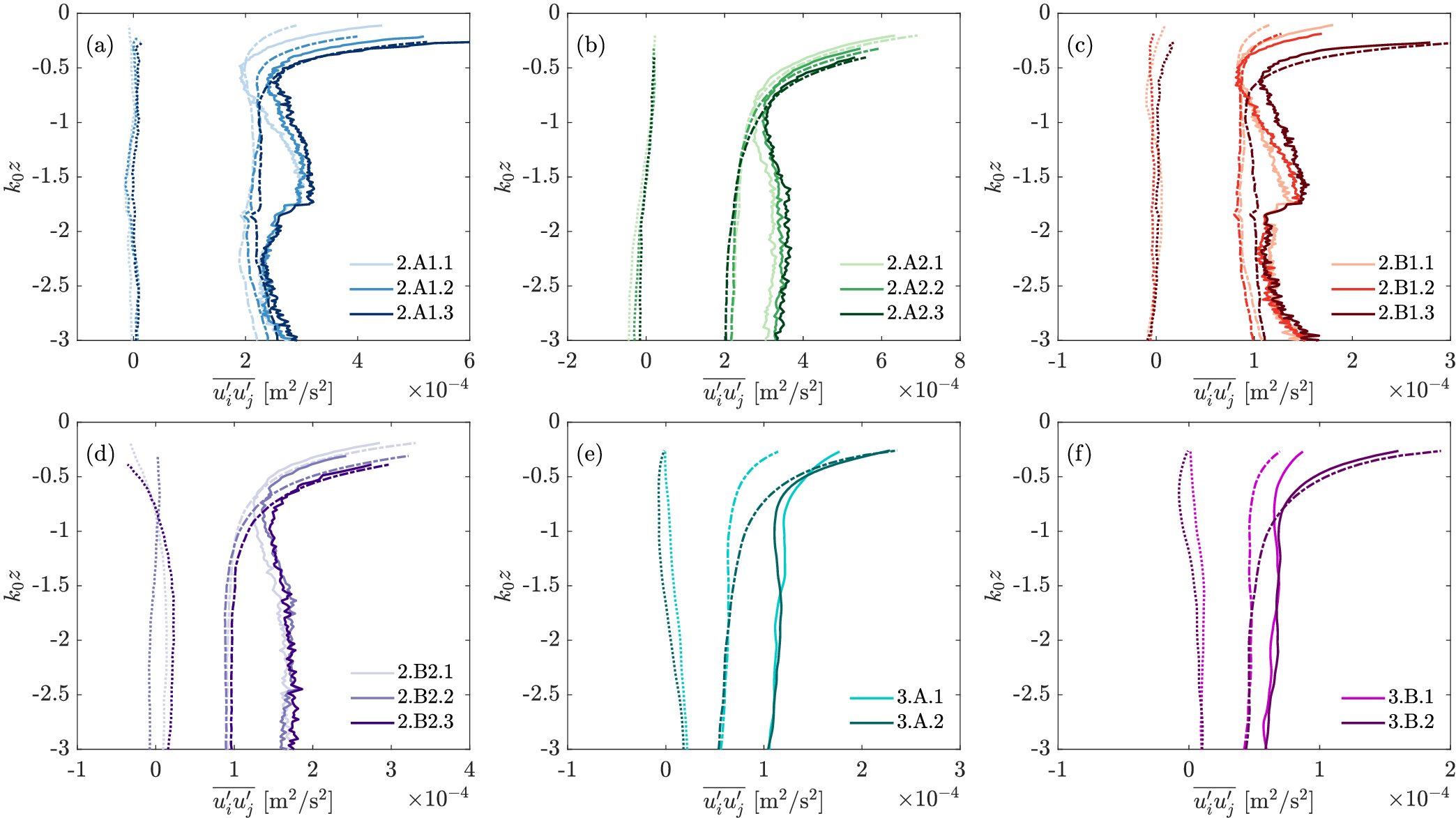} \\
    \includegraphics[width=\linewidth]{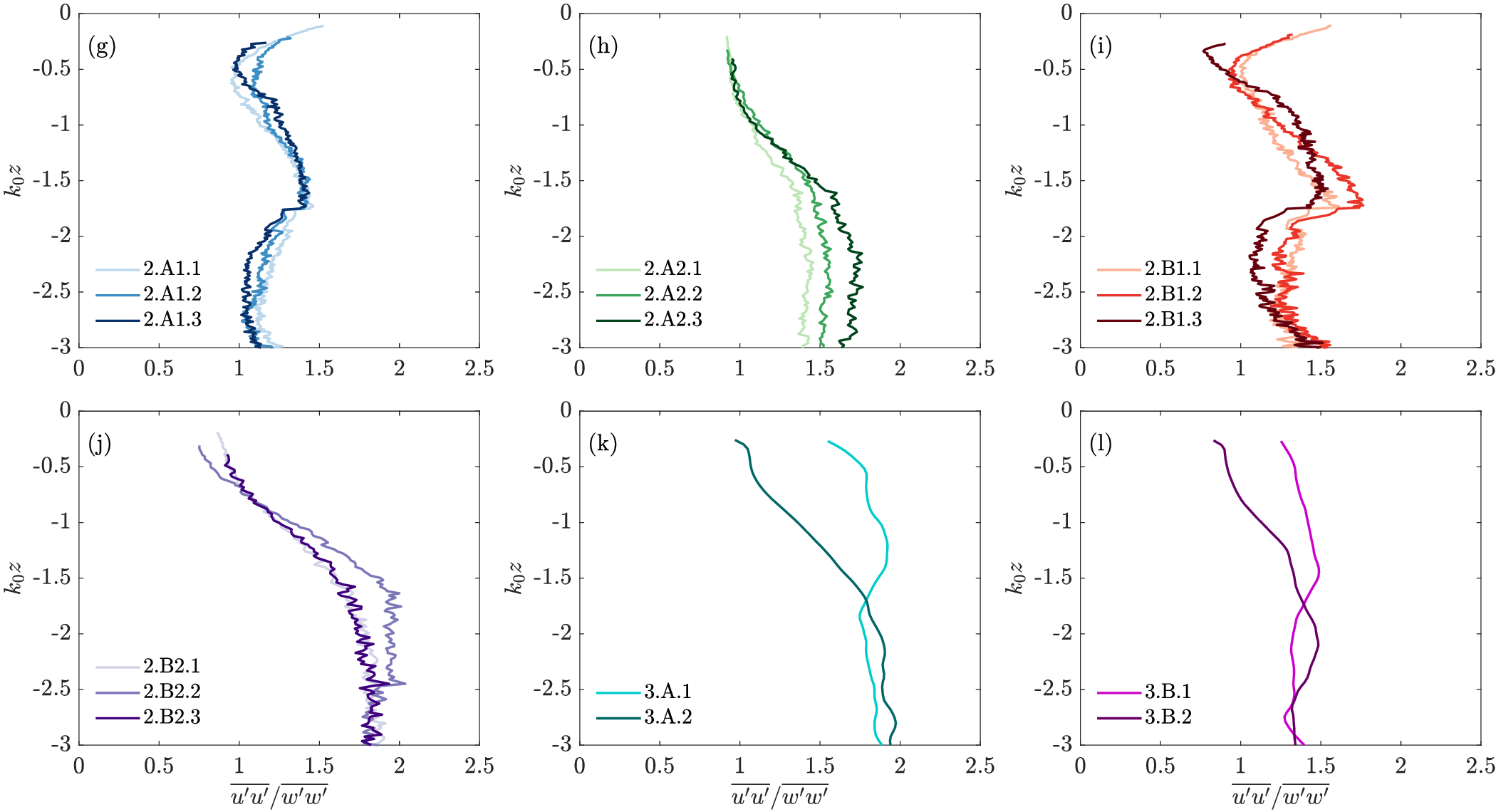} 
    \caption{(a)-(f) Plots of Reynolds stresses $\ouu$ (solid lines), $\oww$ (dash-dotted lines) and $\ouw$ (dotted lines) for all cases. (g)-(l): The ratio $\ouu/\oww$ for all cases. Colours distinguish each case as described in the legends of each panel.}
    \label{fig:Appendix_ReynoldsStress}
\end{figure}

For deeper consideration of our results such as figure \ref{fig:StokesDriftVsEuler_EH}, it is instructive to regard the Reynolds stresses measured for the cases \iE--\iH\ with regular waves. These are shown in figure \ref{fig:Appendix_ReynoldsStress}; panels (a)--(f) show the stresses themselves while the ratio $\ouu/\oww$ which enters in \eqref{eq:EulerStokes} is shown in panels (g)--(l). The vertical behaviours of $\ouu$ and $\oww$ are qualitatively similar to those found numerically by \cite{Fujiwara2020b} (their figure 8; we cannot reliably capture the region very close to the surface where $\oww$ must by necessity tend to zero.)

\bibliographystyle{jfm}

\end{document}